# Terahertz Communications Using Effective-Medium-Slot Waveguides

Nguyen H. Ngo,[1] Weijie Gao,[1] Guillaume Ducournau,[2] Hadjer Nihel Khelil,[2] Rita Younes,[2]
Pascal Szriftgiser,[3] Hidemasa Yamane,[4] Yoshiharu Yamada,[4] Shuichi Murakami,[4]
Withawat Withayachumnankul,[5] and Masayuki Fujita[1]
[1)] *Graduate School of Engineering Science, The University of Osaka, Toyonaka 560-8531, Japan*
[2)] *CNRS, UMR 8520 - IEMN, Institut d'Electronique de Microélectronique et de Nanotechnologie, Université de Lille, Lille 59000, France*
[3)] *Laboratoire de Physique des Lasers, Atomes et Molécules, UMR CNRS 8523 PhLAM, Université de Lille, Lille 59000, France*
[4)] *Osaka Research Institute of Industrial Science and Technology, Izumi 594-1157, Japan*
[5)] *Terahertz Engineering Laboratory, Adelaide University, Adelaide, SA 5005, Australia*

(*Electronic mail: ngo.h-nguyen.es@osaka-u.ac.jp (N.N.), gao.weijie.es@osaka-u.ac.jp (W.G.), fujita@ee.es.osaka-u.ac.jp (M.F.))

All-dielectric effective-medium-clad waveguides have been widely exploited in terahertz communications owing to their extremely low loss, low dispersion, and broad bandwidth. Here, we propose a substrateless effective-medium-slot waveguide. Additionally, we introduce a taper-free interface that allows terahertz waves to directly couple from a metallic hollow waveguide without requiring dielectric insertion. We achieve a 3-dB bandwidth of 40% with a maximum coupling efficiency of 90% in the WR-2.2 band (330–500 GHz). By employing a broadband uni-traveling-carrier photodiode transmitter and sub-harmonic mixer receivers, we demonstrate an aggregated data rate of 0.8 Tbit/s with quadrature amplitude modulation schemes across 14 channels from 330–600 GHz. The effective-medium-slot waveguide platform yields robust broadband coupling with enhanced mechanical protection, offering reliable interconnects for ultra-high-speed terahertz integrated systems.

## I. INTRODUCTION

In recent years, the rapid expansion in internet services has resulted in an exponential increase in data transfer volumes, with peak wireless data rates expected to exceed 100 Gbit/s in commercial systems by 2030.[1–4] Based on Shannon's theory[5], broader bandwidths can offer higher data capacity. Then, the growing demand for higher data rates is driving the adoption of terahertz frequencies in communications technologies (6G and beyond) owing to their broad spectra.[6,7] With the availability of terahertz spectra around 300-GHz band,[8,9] data rates exceeding 100 Gbit/s were achieved.[10–12] More recently, an aggregated data rate over 1 Tbit/s was achieved in the 600-GHz band using a system of a photonic-based transmitter (Tx) with an electronic receiver across 14 channels.[13] However, this demonstration was limited to a back-to-back configuration, in which the Tx and receiver (Rx) were directly connected. The realization of higher data rates at frequencies above 300 GHz remains challenging due to the limited availability of high-performance terahertz interconnects.

Terahertz waves can be generated using several techniques, such as down-conversion of optical signals,[14] frequency multiplication,[15–17] and self-oscillation of semiconductor devices.[18,19] The generated terahertz waves usually need to be coupled to various interconnects, such as metallic hollow waveguides, fibers, and dielectric waveguides, which can operate up to 1 THz or higher.[20] Notably, metallic hollow waveguides, in which most of the waves propagates in the air core, are widely used in commercial devices and lab demonstrations.[21] With advanced micromachining processes, the size of waveguides can be as small as 43 µm for operation frequencies up to 3 THz.[22–24] However, the propagation loss in metallic waveguides, even with a gold-plating layer, becomes significant at such high frequencies due to dramatically reduced skin depths.[25] Although metallic waveguides are employed as a de facto standard interconnect in terahertz system, their flange design, which is usually fixed by standard rather than operation frequency. This makes them relatively bulky, occupying considerable space and limiting their applicability in large-scale integration.

As an alternative for on-chip integration, all-dielectric waveguides have proven effectiveness and scalability for terahertz applications. To eliminate substrate-induced dielectric loss and parasitic modes inherent to conventional platforms such as silicon-on-glass,[26] substrateless waveguide platforms were employed to enable intrinsically low-loss and well-confined terahertz waves.[25,27–30] In particular, effective-medium (EM)-clad waveguides made of high-resistivity intrinsic silicon exhibited extremely high efficiency, low dispersion, and broad bandwidth.[28] They offer propagation losses as low as 0.05 dB/cm and a fractional bandwidth as large as 42%. Importantly, the EM cladding formed by a periodic array of air holes can be modeled as a homogeneous material with tailorable permittivity tensor,[28,31] resulting in the realization of various components for integrated terahertz devices,[32–37] as illustrated in Fig. 1. In conventional designs, tapered spikes with a length of several wavelength are employed to mitigate the mismatch caused by the large refractive index contrast between silicon and air at the hollow waveguide's opening[27]. As the operating frequency approaches the WR-2.2 band (330–500 GHz) and beyond, the opening of the hollow core is 559 µm × 279 µm or smaller, causing difficulties in the waveguide alignment and potentially leading to undesired standing waves and degraded transmittance. Thus, a robust coupling structure is essential to reduce the undesirable effects when coupling silicon waveguides to metallic waveguide-based components.

As an alternative to a solid silicon core, dielectric slot waveguides have attracted interest for their strong field confinement in a low-index region, low propagation loss, and simple fabrication.[38–42] The fundamentals of dielectric slot waveguides were investigated in both optical[38,43] and terahertz domains.[44] At high-index-contrast interfaces, continuity of the normal component of the electric displacement field enforces a strong electric-field ($E$-field) discontinuity, resulting in significant field enhancement and confinement of the guided mode within the low-index air gap. Owing to the strong field localization in the low-index region, slot waveguides are attractive for sensing applications. This enhanced light-matter interaction has enabled highly sensitive refractive-index sensing, biochemical detection, and imaging platforms.[45] In addition, the slot concept is applied in various passive components, such as antenna,[44,46] coupler,[47–49] and waveguides.[36,50] Earlier, we first demonstrated non-tapered-spike coupling by using a modified slot to minimize index mismatch at the interface with a metallic waveguide[51], achieving 32-Gbit/s error-free transmission by using on-off keying modulation. Nevertheless, an additional structure, i.e., photonic crystal slab, was required to enhance the coupling efficiency at low frequencies, which increased the design complexity.

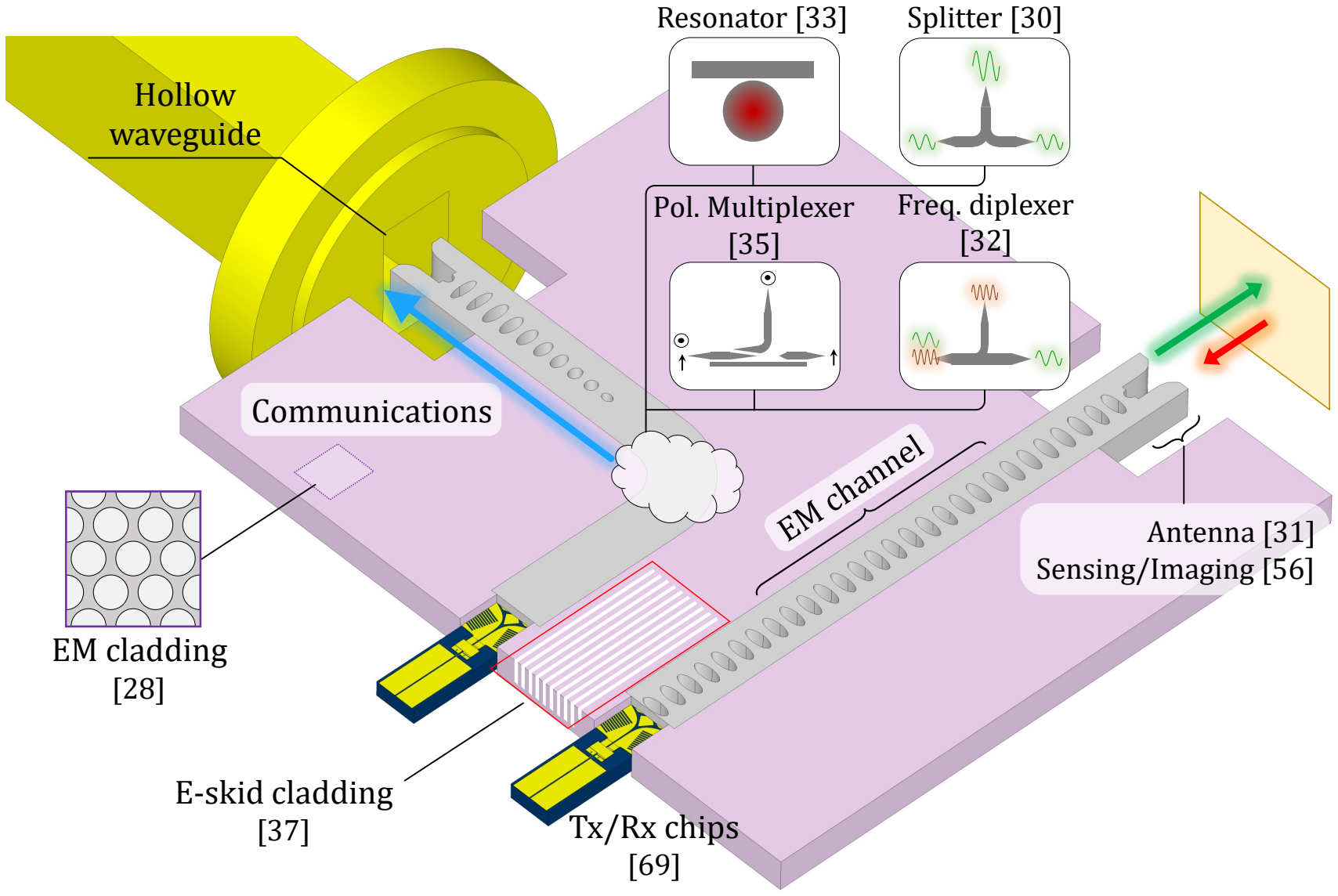

FIG. 1. Integrated all-silicon platform comprising EM-slot waveguides. Active devices are coupled to the EM channel for sensing and communication applications.

In this work, we introduce a substrateless slot waveguide with an EM structure incorporating a spikeless coupler, featuring robust coupling between all-silicon slot waveguide and hollow waveguide in the WR-2.2 band (330–500 GHz). The slot coupler is optimized to reduce mode mismatch when connecting to the hollow waveguide, enhancing the coupling efficiency up to 90%. To realize the slot waveguide without a supporting substrate, the slot is substituted by an array of sub-wavelength elliptical air hole. This silicon-air structure forms an anisotropic low-index region, efficiently transferring the energy captured from the slot coupler. The developed substrateless slot waveguide is employed as a terahertz interconnect capable of operating beyond 500 GHz, and enabling a broad bandwidth for communications with aggregated data rates of up to 809 Gbit/s, thereby demonstrating its potential for ultra-high-speed data transmission. By incorporating compact terahertz transceivers,[52,53] the waveguide architecture can be expected to support non-destructive sensing and imaging,[54–56] facilitating integrated sensing and communication applications for 6G and beyond, as illustrated in Fig. 1.

Section II describes the design and analyses of the substrateless slot waveguide. Section III presents the characterization of the fabricated samples. Section IV demonstrates the broad bandwidth capability by supporting communications rates of 0.8 Tbit/s. Section V compares the EM-slot waveguide with other existing waveguides in the terahertz range and discusses potential improvements to the waveguide design.

## II. REVOLUTION OF THE WAVEGUIDE DESIGN

In practice, a conventional slot waveguide consisting of an air slot is not mechanically feasible for the substrateless platform.[44] Therefore, we implement a slot-based EM channel, where the electromagnetic field remains confined within a low-index region with enhanced mechanical stability. Especially, this EM channel also enables a tailorable effective refractive index for improved matching and functional purposes (e.g., sensor, coupler). The underlying design principle of the proposed waveguide is the efficient matching of mode profiles at the interface of the slot waveguide to a target structure, i.e., a metallic hollow waveguide. The mode supported by the slot coupler should closely resemble the fundamental transverse-electric (TE) mode of the hollow waveguide to maximize spatial field overlap. At the same time, the effective refractive index should be properly matched to suppress the mismatch arisen from index discontinuities at the interface.

The proposed substrateless slot waveguide is conceptually divided into two functional sections: (i) an EM channel that forms a low-index material of the slot and offers mechanical support for the dual silicon beams, and (ii) a slot coupler for taper-free coupling between the guiding channel and a metallic hollow waveguide. The design target is to maximize the coupling efficiency between the slot waveguide and the metallic hollow waveguide while keeping a broad 3-dB bandwidth. All structures are simulated using a full-wave simulator, CST Microwave Studio Suite. In each configuration, the dielectric waveguides are fed by metallic hollow waveguides via contact interfacing as a practical implementation. The operation frequency covers the WR-2.2 band. The design concepts are detailed as follows.

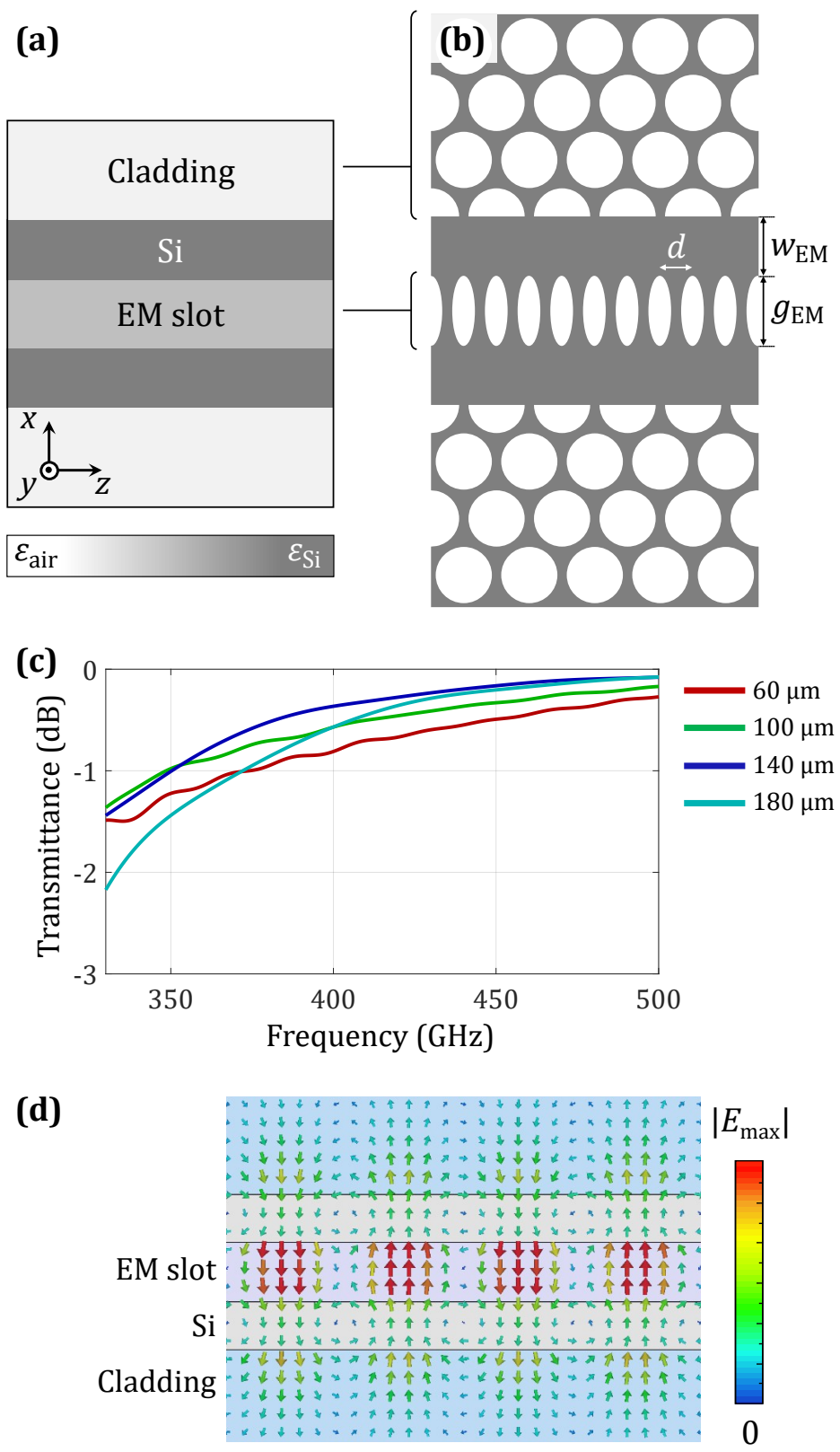

FIG. 2. EM-slot waveguide. (a) Representation of slot and cladding replaced by artificial anisotropic materials. (b) EM-slot waveguide based on air-silicon (Si) periodic structures. (c) Transmittance of EM channel with various channel gaps $g_{\mathrm{EM}}$. (d) Simulated instantaneous $E$-field vector distribution at 415 GHz.

### A. Effective-medium-slot waveguide

A bare slot waveguide is typically fabricated on a silicon-on-insulator platform. However, this configuration introduces additional absorption and radiation losses due to the wave leakage into the substrate. Moreover, the need for adhesive bonding[47] or heterogeneous material integration approach[42] further increases fabrication complexity.

To this end, we propose a substrateless slot waveguide, as conceptualized in Fig. 2(a). A fundamental TE mode with in-plane polarization is excited, where the $E$-field component $E_x$ propagates along the $z$-direction. A central EM channel with a gap $g_{\mathrm{EM}}$ acts as a low-index region to connect the two silicon beams with a width $w_{\mathrm{EM}}$, while additional EM claddings with refractive indices close to that of air are applied to the outer sides of the beams for wave confinement and handling purposes.[28]

To realize an all-silicon substrateless structure, an effective medium in the form of elliptical air holes is introduced into the low-index region, as shown in Fig. 2(b). This air hole array also establishes a connection for the two silicon beams. Since the structure contains several design parameters, we first determine the dimensions of an equivalent homogeneous EM channel characterized by anisotropic permittivity. The initial conditions for the total channel width $(2w_{\mathrm{EM}} + g_{\mathrm{EM}})$ are fixed at 260 μm, which is close to the width of the core of the metallic hollow waveguide (279 μm). Elliptical air holes are designed with a period $d$ of 50 μm, which is much shorter than a quarter of the wavelength of 500 GHz in bulk silicon. The gap between two elliptical air holes should be larger than 10 μm, which is limited by fabrication to avoid over-etching. This gives the semi-minor axis $(r_a)$ to be 20 μm. This subwavelength-featured channel is therefore considered as an anisotropic material, where the relative permittivity tensor of the EM slot $(\varepsilon_{\mathrm{x}}, \varepsilon_{\mathrm{y}}, \varepsilon_{\mathrm{z}}) = (3.65, 5.04, 3.65)$ is derived as follows[28,36]:

$$\varepsilon_{\mathrm{x}} = \varepsilon_{\mathrm{z}} = \varepsilon_{\mathrm{Si}} \frac{(\varepsilon_{\mathrm{Si}} + \varepsilon_0) + (\varepsilon_0 - \varepsilon_{\mathrm{Si}})\zeta}{(\varepsilon_{\mathrm{Si}} + \varepsilon_0) - (\varepsilon_0 - \varepsilon_{\mathrm{Si}})\zeta}, \tag{1}$$

$$\varepsilon_{\mathrm{y}} = \varepsilon_{\mathrm{Si}} + (\varepsilon_0 - \varepsilon_{\mathrm{Si}})\zeta, \tag{2}$$

$$\zeta = \frac{\pi r_{\mathrm{a1}} r_{\mathrm{b1}}}{2 r_{\mathrm{b1}} d}, \tag{3}$$

where $\varepsilon_{\mathrm{Si}} = 11.68$ and $\varepsilon_0 = 1$ represent the relative permittivities of silicon and air, and $\zeta$ is the filling factor of air volume in silicon. In order to avoid under-etching of the air holes, the minimum $g_{\mathrm{EM}}$ is set to 40 μm. The gap of the EM-channel $(g_{\mathrm{EM}})$ is varied, and the transmittance is shown in Fig. 2(c). It is noted that the transmission at 180 μm reduces at the low-frequency range. As the gap increases, the width of silicon beam reduces. The mode profile at this low-frequency range is wider and causes leakage to the air cladding. The highest average transmittance is used to evaluate the waveguide performance, which is obtained when the gap is 140 μm. As a result, initial vales of channel width $(w_{\mathrm{EM}})$ and gap $(g_{\mathrm{EM}})$ are set to 60 μm and 140 μm, respectively. Similarly, the relative permittivity tensor of the cladding, of which the period $a_{\mathrm{clad}}$ is 120 μm and radius $r$ is 55 μm, is calculated as $(\varepsilon_{\mathrm{x}}, \varepsilon_{\mathrm{y}}, \varepsilon_{\mathrm{z}}) = (2.55, 3.54, 2.55)$. The relative permittivity tensors of the EM slot and cladding are employed in the simulation model built from Fig. 2(a), showing that the guided mode is confined within the EM slot region, as illustrated in Fig. 2(d). This approach effectively realizes a slot waveguide without relying on a conventional supporting substrate. By eliminating substrate-induced dielectric loss, the proposed structure enables high efficiencies for terahertz-wave propagation.

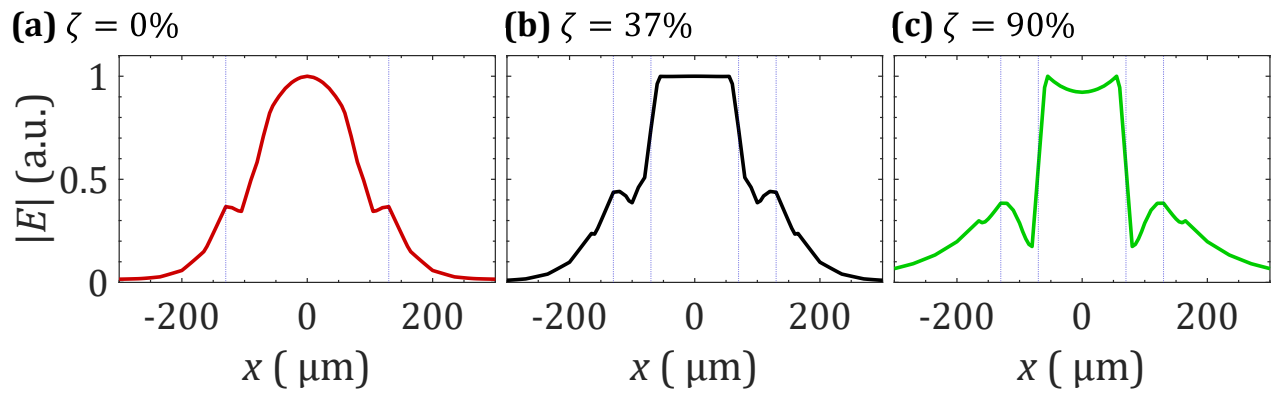

FIG. 3. $E$-field distribution along $x$-axis, normalized to its maximum, at different filling factor $\zeta$: (a) 0% (corresponding to a solid waveguide), (b) 37 %, (c) 90 %. Channel width $w_{\mathrm{EM}} = 60$ μm and gap $g_{\mathrm{EM}} = 140$ μm.

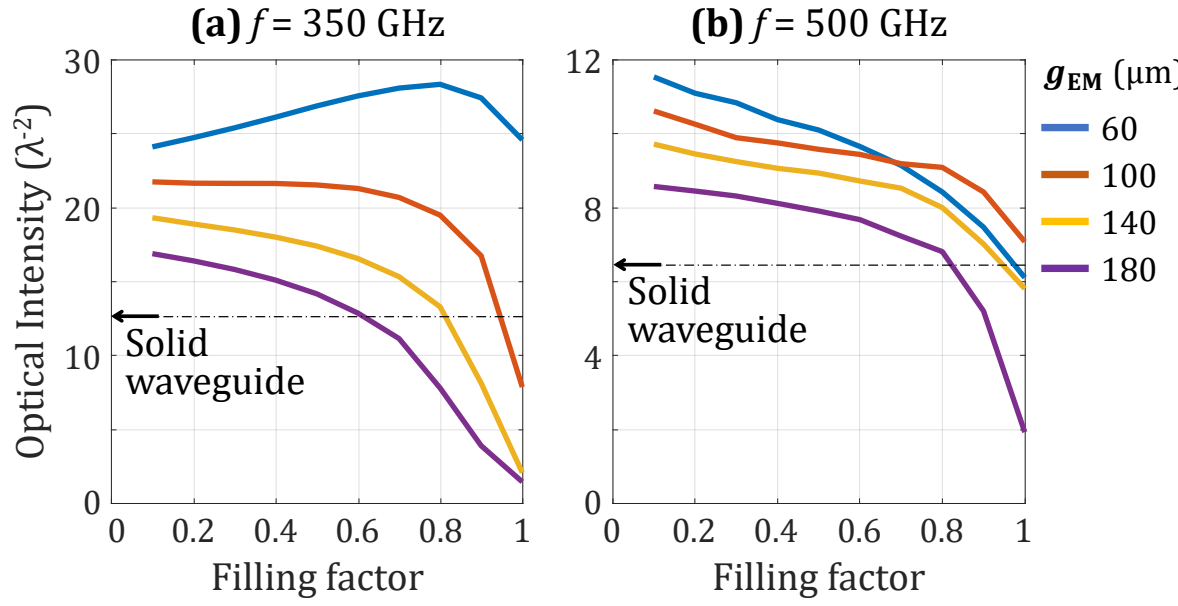

FIG. 4. Optical intensity $I$ of the EM slot waveguide with various channel gap $g_{\mathrm{EM}}$ at (a) 350 GHz and (b) 500 GHz. Optical intensity of solid waveguide is also indicated.

Compared with conventional EM-clad and air-slot waveguides, the EM-slot waveguide provides an additional degree of freedom to modify the modal index of the dielectric waveguide without relying on additional materials. Specifically, the EM-slot configuration offers greater design flexibility to tailor field confinement by adjusting the filling factor through an air-hole lattice. Figure 3 illustrates the ability to control the $E$-field distribution by tuning the relative permittivity tensor of the EM channel with different filling factors $\zeta$. The underlying mechanism is that the effective index determines the evanescent-field decay inside the slot, thereby modifying the $E$-field profile.[42] For instance, when the filling factor is $\zeta = 37\%$, the $E$-field inside the slot becomes uniformly distributed because the field decay coefficient $k$ in the slot approaches zero. This uniform field distribution is independent on the slot gap, as demonstrated in Mach–Zehnder interferometers designed as electro-optic modulators.[42] For light-matter interaction applications,[45] engineering the relative permittivity closer to that of air (e.g., $\zeta = 90\%$) enables strong $E$-field localization in the low-index region.

Additionally, the optical intensity in the slot can be expressed as $I_{\mathrm{EM\ slot}} = P_{\mathrm{EM\ slot}}/(hg_{\mathrm{EM}})$ as a function of wavelength, where $P_{\mathrm{EM\ slot}}$ represents the power confined in the slot normalized to the total waveguide power,[38] with the result shown in Fig. 4. For comparison, the normalized intensity of a solid silicon waveguide with the same total cross-sectional area ($hW_{\mathrm{total}}$) is also included. At both frequencies, the optimized EM-slot waveguide (with $\zeta = 0.62$) exhibits stronger field confinement than the solid silicon waveguide. In on-chip communications, suppressing inter-channel interference is critical for densely integrated waveguide arrays. In the EM-slot waveguide, the evanescent decay inside the slot can be engineered by adjusting the filling factor of the air holes, thereby modifying the effective permittivity tensor and the corresponding transverse wavevector,[57] so that the optical energy is concentrated inside this low-index area. Additional extreme skin-depth (e-skid) claddings[37] can further suppress evanescent fields outside the core, thereby reducing modal overlap between adjacent channels, showing the potential for terahertz integrated platforms.

## B. Slot coupler

Metallic hollow waveguides are commonly used to excite terahertz dielectric waveguides. Figure 5(a) shows an in-plane cutaway view of the EM-slot waveguide, of which design parameters are obtained in Sec. II.A, coupled to the hollow waveguide. The abrupt transition at the interface between the air at the hollow waveguide output and the silicon of the slot waveguide causes large mode mismatch, leading to significant fluctuations in transmission and reflection, as shown in Fig. 5(d). In this section, the coupling structure between a WR-2.2 metallic hollow waveguide and the EM-slot waveguide is designed to improve coupling efficiency. First, the initialization for the parameters of the slot coupler is determined to maximize the coupling efficiency ($S_{21}$) and minimize the reflectance ($S_{11}$). Then, the global optimization using a built-in optimizer in CST Studio Suite is applied for the final design. Details of the optimization are provided in *Supplementary Material–S1*.

Specifically, a slot coupler consisting of two trapezoidal beams is first introduced while the channel remains homogeneous silicon for matching from the hollow waveguide to the EM channel, as shown in Fig. 6(a). The slot parameters with a gap $g$ and length $L_{\mathrm{slot}}$ are systematically swept to provide averages transmittance and reflectance in the whole WR-2.2 band. The optimum $S$-parameter, that is the lowest average $|S_{11}|$ and highest average $|S_{21}|$, is achieved when $g$ and $L_{\mathrm{slot}}$ are 140 μm and 150 μm, as shown in Fig. 6(b,c). Next, the homogeneous channel is substituted by the EM structure consisting of an array of elliptical air holes, as shown in Fig. 6(d). However, this introduction of air holes can cause reflection at the interface. To enable a smooth modal transition between the coupler and the EM waveguide, the slot profile is apodized using an $n^{th}$-order polynomial taper with an offset $\Delta s$. By using the same method for sweeping the curvature and offset, the modal evolution along the transition becomes smoother compared to the EM-slot channel without the couplers, leading to lower reflection across the WR-2.2 band, as shown in Fig. 5(e).

Although the reflection decreases, the average transmittance is approximately –3 dB, which is attributed to the absence of a transition between the slot coupler and the edge of the air-hole array, causing slightly fluctuating $S$-parameters. To further improve the mode matching, the first air hole at the interface is modified as a matching element, as depicted in Fig. 5(c). This configuration provides an appropriate initial condition for global optimization. Finally, a covariance matrix adaptation evolution strategy optimizer is applied to re-fine both parameters of the slot couplers and the EM channel. The resulting S-parameters demonstrate enhanced transmission and reduced reflection across the band, as presented in Fig. 5(f), and the corresponding optimized structure is shown in Fig. 5(g).

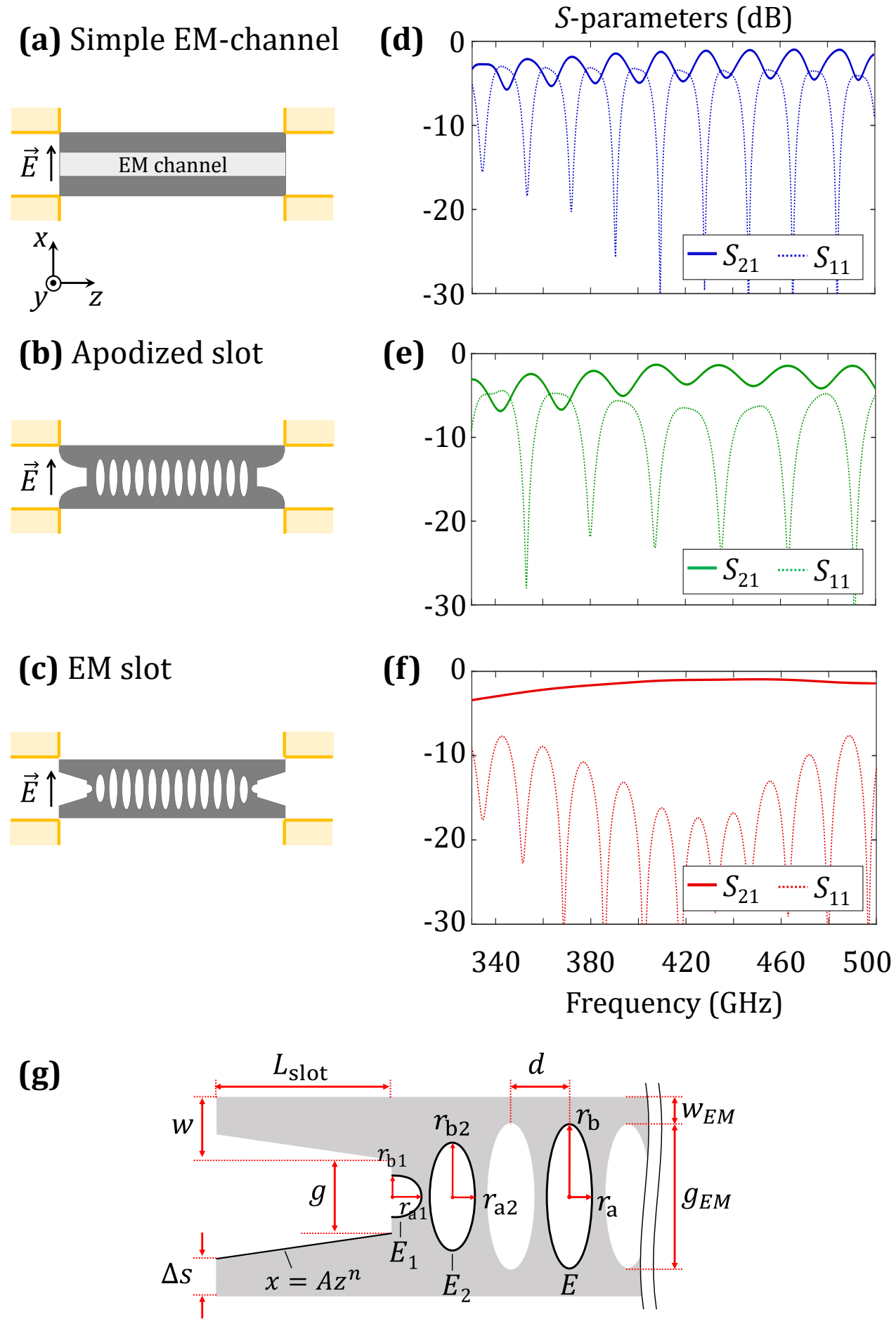

FIG. 5. In-plane cutaway views of slot-based silicon waveguides (gray) coupling to hollow rectangular waveguides (yellow): (a) Simple EM channel, (b) apodized slot, and (c) EM-slot waveguide. $S$-parameters of (d) simple EM-channel waveguide, (e) apodized-slot waveguide, and (f) EM-slot waveguide. (g) Magnified view of the EM-slot waveguide. Parameters of are given as: $w$ = 80 μm, $g$ = 80 μm, $l_{slot}$ = 145 μm, $n$ = 1, $\Delta s$ = 70 μm; $d$ = 50 μm, $r_a$ = 19.8 μm, $r_b$ = 90 μm, $r_{a1}$ = 26.4 μm, $r_{b1}$ = 30 μm, $r_{a2}$ = 19.8 μm, $r_{b2}$ = 72 μm.

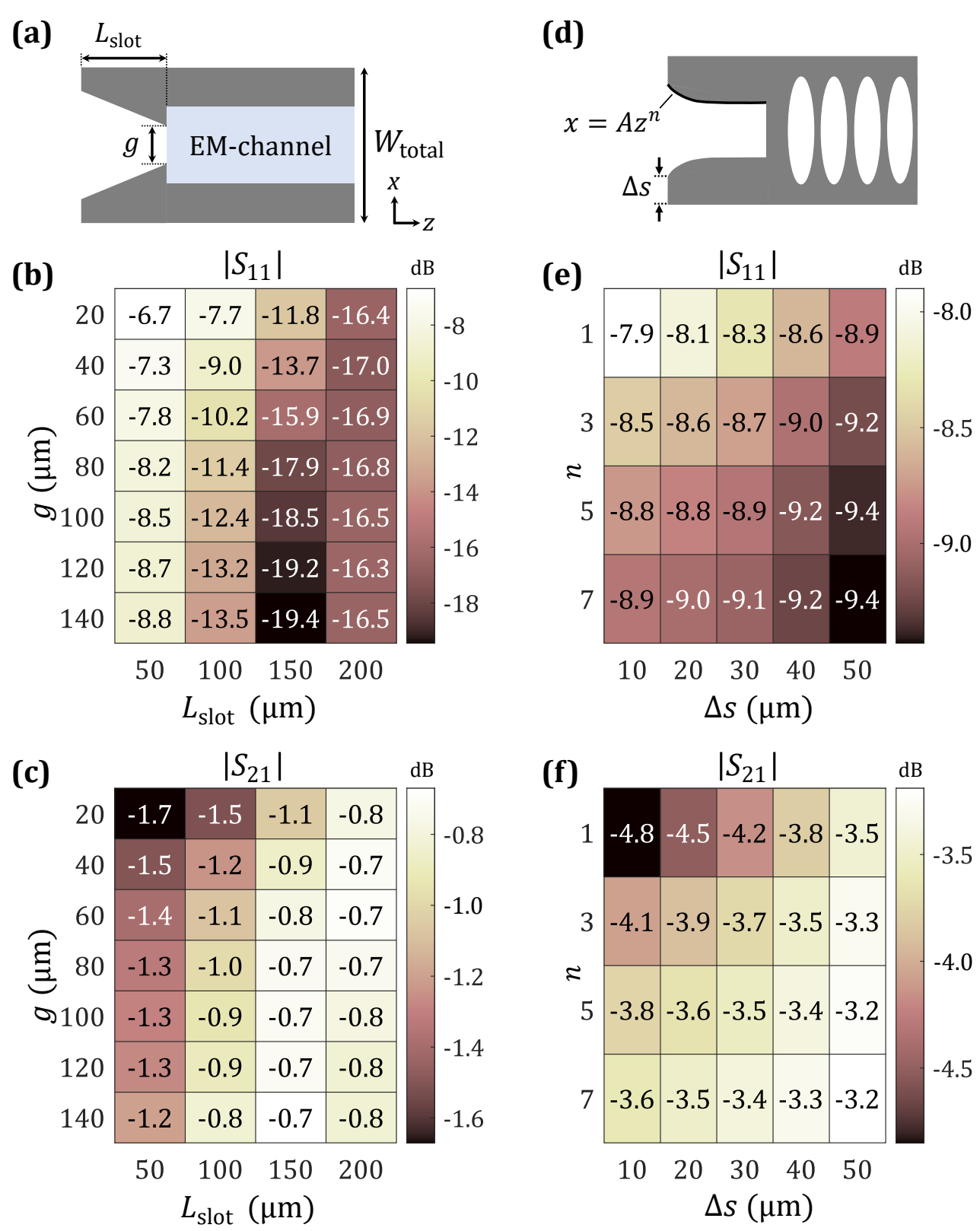

FIG. 6. Parameter optimization of the slot couplers with heatmap of average $S$-parameters. Slot coupler incorporated into homogeneous EM channel with (a) schematic, (b) $|S_{11}|$and (c) $|S_{21}|$. Slot coupler incorporated into elliptical air-hole channel with (d) schematic, (e) $|S_{11}|$, and (f) $|S_{21}|$.

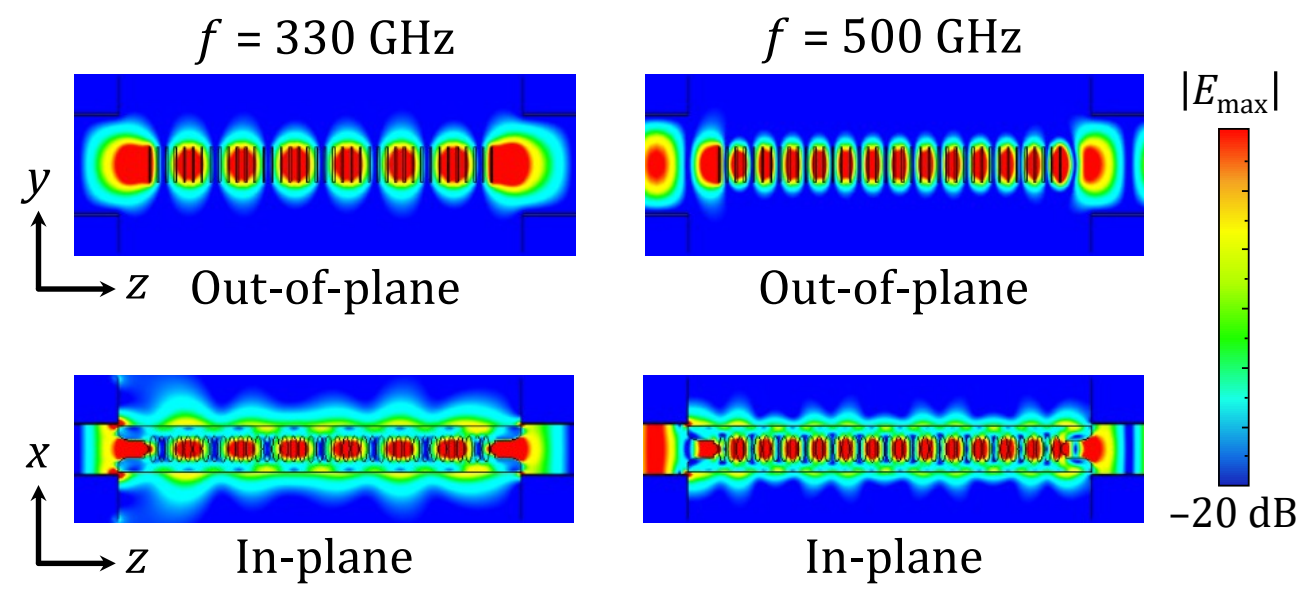

FIG. 7. Field plots for the proposed waveguide. Out-of-plane and in-plane 2D $E$-field distributions of the substrateless slot waveguide at 330 GHz and 500 GHz.

The simulated $E$-field distributions in the substrateless slot waveguide at different frequencies are illustrated in Fig. 7. In the in-plane cross-section, the waves can efficiently couple into the slot and propagate along the elliptical air holes, where the guided modes are confined between the dual silicon beams. At high frequencies (e.g., 500 GHz), the guided mode exhibits stronger confinement, resulting in high coupling efficiency of approximately −1 dB. In contrast, at the lower frequency bound, the weaker wave confinement arises from the longer wavelength, which prevents the slot from capturing the entire field energy. This effect, combined with the limited thickness of the 200-μm silicon wafer, leads to wave leakage and scattering into air, explaining the reduced coupling efficiency of −3.5 dB at 330 GHz, as shown in Fig. 5(f). Despite the slight degradation at lower frequencies, the simulated $S$-parameters suggest a wide 3-dB bandwidth of 170 GHz (41% fractional bandwidth) and an average coupling efficiency of −1.5 dB, which are sufficient for most terahertz applications such as high-speed communications. More importantly, the proposed design improves practical implementation by eliminating the fragile tapered spikes traditionally required for coupling with metallic hollow waveguides.

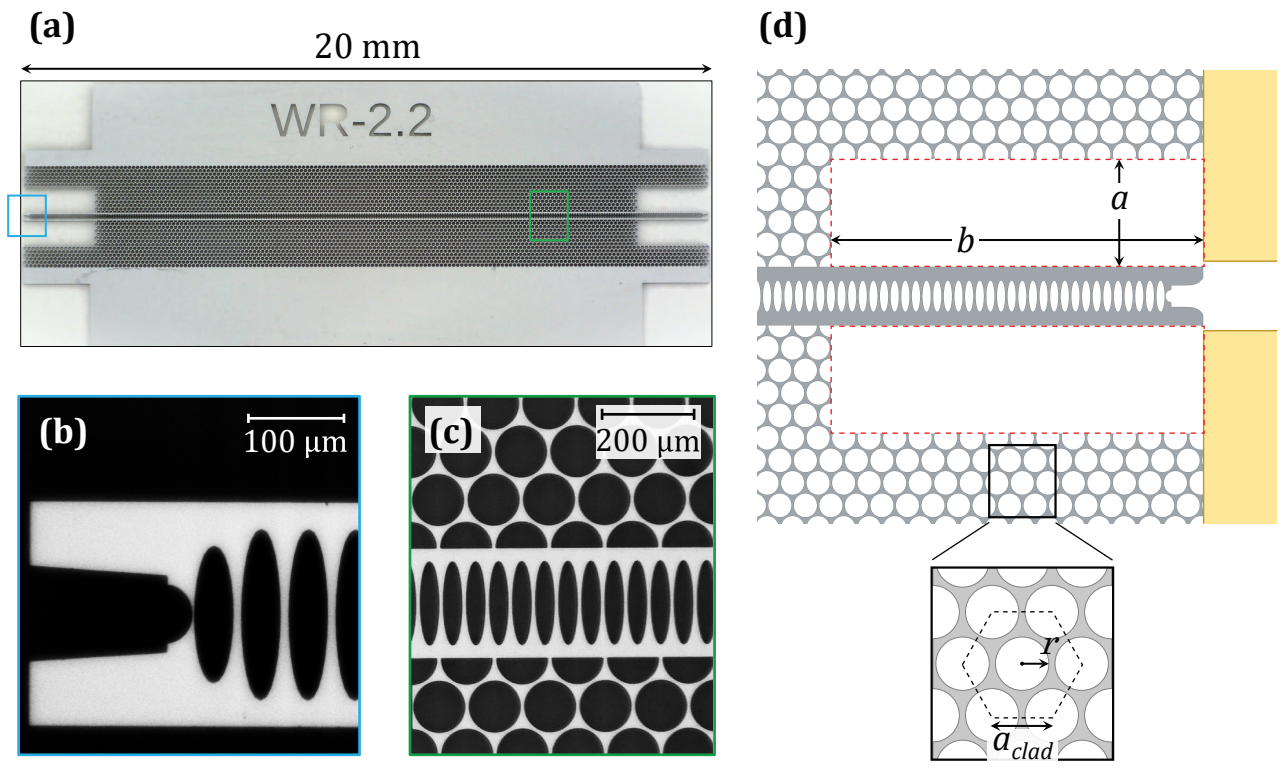

FIG. 8. Fabricated substrateless slot waveguide. (a) Fabricated sample with length of 20 mm. Magnified views of (b) slot coupler and (c) EM slot, consisting of elliptical air holes. (d) Dimensions of EM claddings, where $a_{\text{clad}} = 120$ µm and $r = 55$ µm. The red-dashed rectangles are trimmed from claddings to reduce the interference as waves couple from the hollow waveguide (yellow) to the slot. The dimensions are $a = 500$ µm and $b = 2650$ µm.

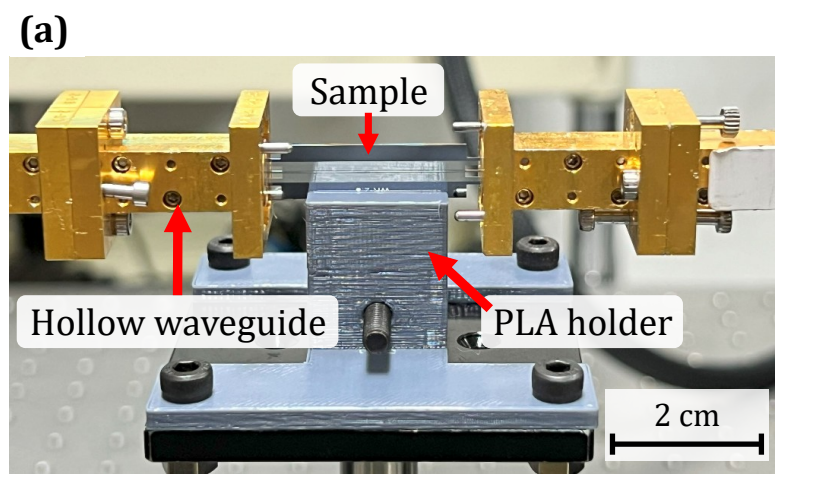

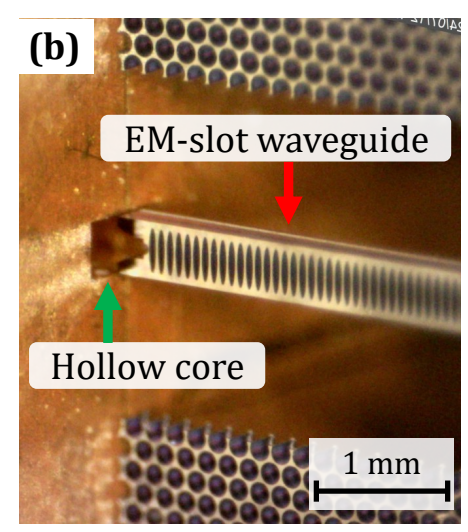

FIG. 9. Experimental setup for characterization. (a) Setup for the S-parameters measurement. (b) Magnified view at the interface between the slot coupler and the metallic hollow waveguide.

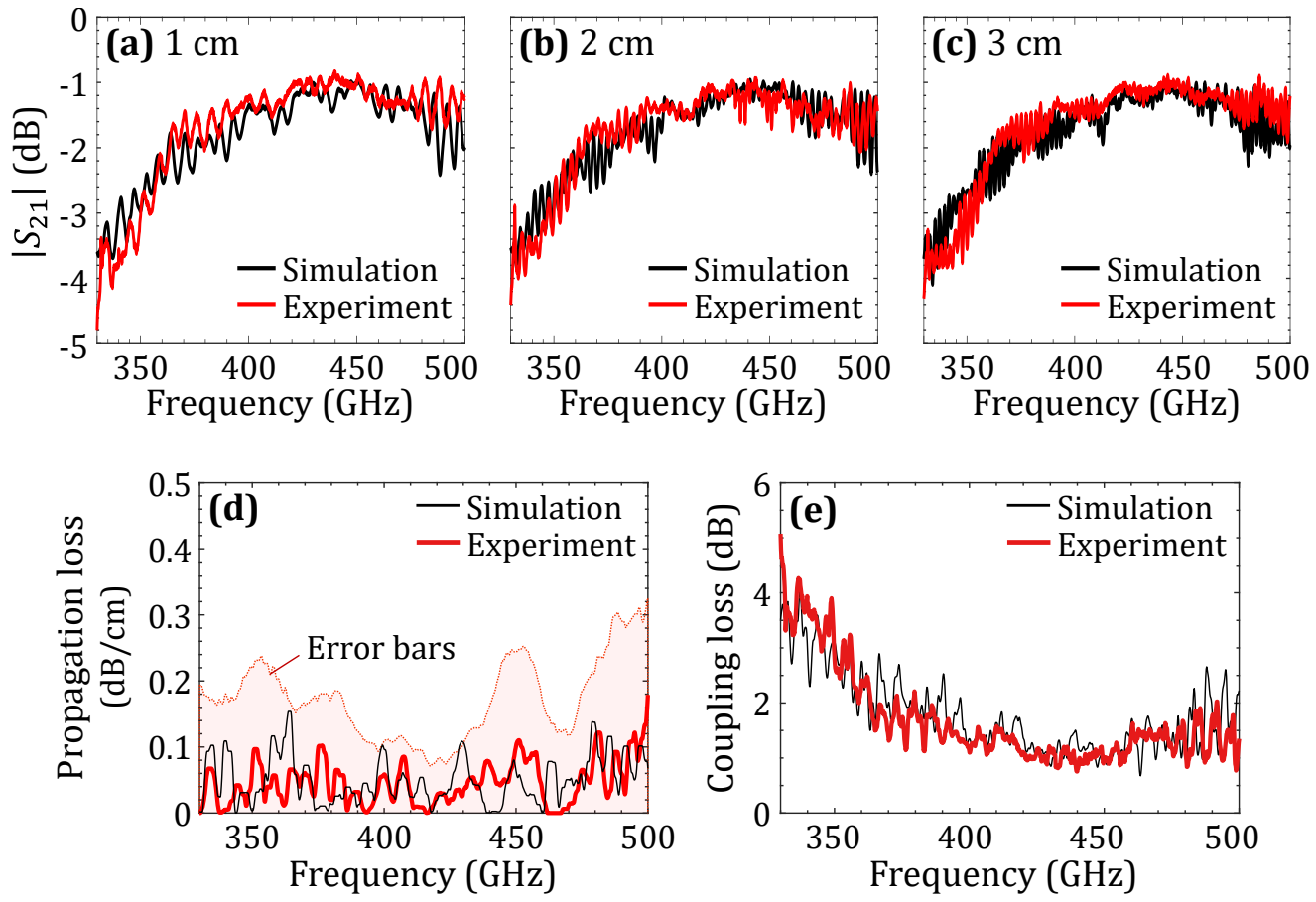

FIG. 10. Characteristics of EM-slot waveguide. Simulated and measured transmission coefficients at lengths of (a) 1 cm, (b ) 2 cm, and (c) 3 cm. (d) Propagation loss with error bars. (e) Coupling loss.

## III. WAVEGUIDE CHARACTERIZATION

The waveguides with different lengths (1, 2, and 3 cm) are fabricated using the deep reactive ion etching (DRIE) process on a high-resistivity (>10 kΩ-cm) float-zone silicon wafer with a thickness of 200 µm. In Figs. 8 (a-c), the fabrication quality is inspected using a microscope, revealing a tolerance of approximately 5% owing to its high precision. To enable the proposed channel waveguide as an integrated platform for accommodating both passive and active components, we introduce EM claddings surrounding the core, as shown in Fig. 8(d). Noticeably, the cladding is trimmed with a rectangular area so that the interference of the evanescent fields at the slot couplers is minimized. The vertical distance $a$ between the slot and the cladding is 500 µm, where the $E$-field intensity is negligible.

### A. Measurements and Results

The S-parameters are measured using a Keysight vector network analyzer (VNA) with VDI extenders operating at the WR-2.2 band. As shown in Fig. 9(a), the waveguide sample is secured using a 3D-printed holder and is directly in contact with the opening of the feeding hollow waveguides, as illustrated in Fig. 9(b). The alignment between the feed and the sample is finely adjusted using micropositioning stages. Figure 10(a-c) shows the coupling efficiencies of the channel waveguides for various lengths. Both the simulation and experimental results exhibit excellent agreements, validating the efficacy of the channel structure. The maximum and average coupling efficiencies of the 3-cm sample are −0.9 and −1.7 dB, respectively. The 3-dB bandwidth spans 165 GHz with a fractional bandwidth of 40%, ranging from 335 GHz to 500 GHz, covers almost the whole WR-2.2 band.

The propagation loss and coupling loss are extracted by comparing the transmission coefficients of the waveguides with various lengths. As shown in Fig. 10(d), the average and maximum propagation losses are 0.05 and 0.38 dB/cm, respectively, which are attributed to the low-loss characteristic of high-resistivity silicon. The presence of negative values in the propagation loss is primarily due to fluctuations in the transmission coefficients caused by coupling between the dielectric and hollow waveguides. As shown in Fig. 10(e), the coupling loss ranges from 0.7 to 5 dB, with an average value of 1.7 dB across the WR-2.2 band. To reduce the coupling loss at a lower frequency range (<350 GHz), a larger aperture of the hollow waveguide, such as WR-2.8 (711 µm × 356 µm), can be used to capture the field leakage, while the corresponding dielectric waveguide interface requires slight optimization to fit the larger aperture.

## IV. BROADBAND WAVEGUIDE FOR HIGH-SPEED COMMUNICATIONS

In the previous section, the measured transmission coefficient showed that the operational bandwidth of the waveguide could extend toward higher frequencies. In order to fully exploit the available broad spectrum for communications, the waveguide is further characterized in the WR-1.5 band (500–

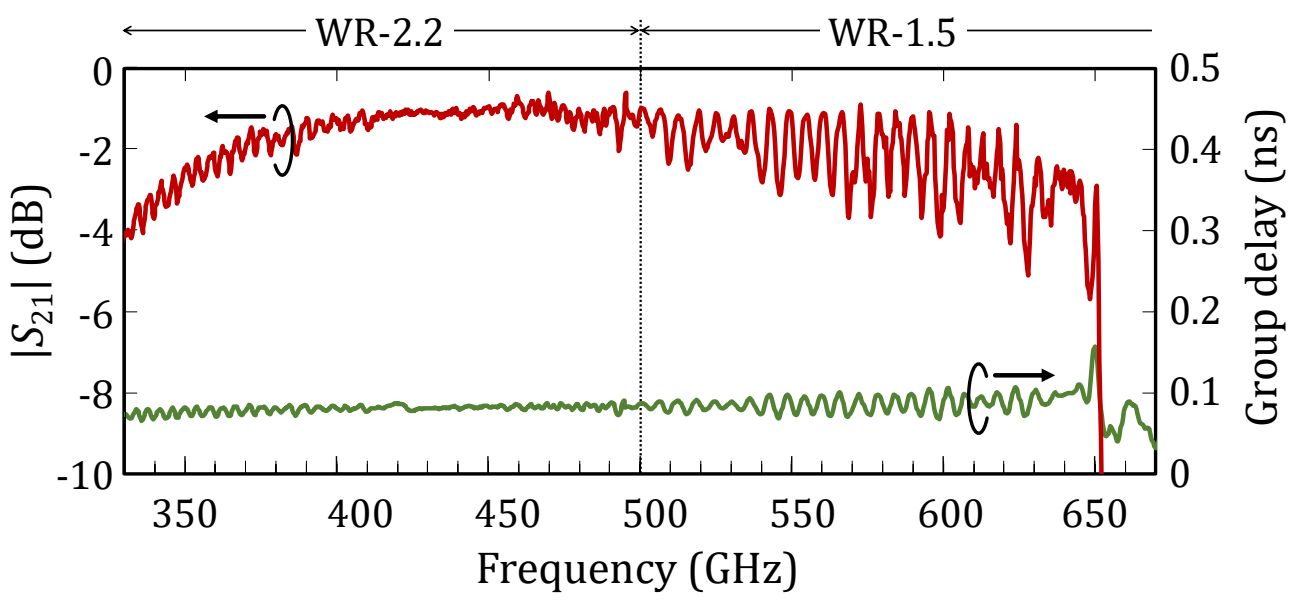

FIG. 11. Measured transmittance and group delay of the substrate-less slot waveguide across the WR-2.2 (330–500 GHz) and WR-1.5 (500–750 GHz) bands.

750 GHz). Figure 11 presents the transmission coefficients and group delay of the channel waveguide across both the WR-2.2 and WR-1.5 bands. Ripples observed in the transmittance at high frequencies arise from the excitation of higher-order modes and increased mode mismatch at the waveguide transitions. As the guided wavelength decreases, the EM channel is no longer in the sub-wavelength regime, leading to increased modal dispersion in the guiding structure. Consequently, the ripples becomes noticeable above 500 GHz with an increase of the group delay. To reduce the ripples, the profiles of the slot coupler and EM-channel can be re-optimized expanding to the whole WR-1.5 band. However, the results in Fig. 11 indicate that the structure remains well matched up to 600 GHz, corresponding to a fractional 3-dB bandwidth of 57% (335–600 GHz). This broadened bandwidth permits the allocation of multiple distinct carrier channels for communications with high-order vector modulation schemes, which is described in detail in the following subsection.

### A. Experimental setup

As shown in Fig. 12, the setup is based on the same concept as the previously reported test system.[13] On the Tx side, two free-running tunable lasers in the C-band (1530–1560 nm) are used to generate an optical beat signal with a frequency difference in the terahertz band. One of the laser sources is vector-modulated by a data signal generated from an arbitrary waveform generator (AWG) using an optical IQ modulator. The modulated laser output is then combined with the other laser via a fiber coupler and subsequently amplified by an erbium-doped fiber amplifier (EDFA). Subsequently, the combined optical signal is photomixed by a uni-traveling carrier photodiode (UTC-PD)[58] to generate a terahertz signal (330–600 GHz), which is then coupled to a WR-2.2 near-field probe (NFP). The NFP is used to feed the silicon waveguide and kept constant for all carrier frequencies.

On the Rx side, a subharmonic mixer (SHM) is used to demodulate the transmitted signal. The SHM is pumped with a local oscillator (LO) close to half of the carrier frequency. This LO signal is produced by a signal generator (SMA100B from Rohde & Schwarz) with a subsequent frequency multiplier. The resulting intermediate frequency (IF) singal, below

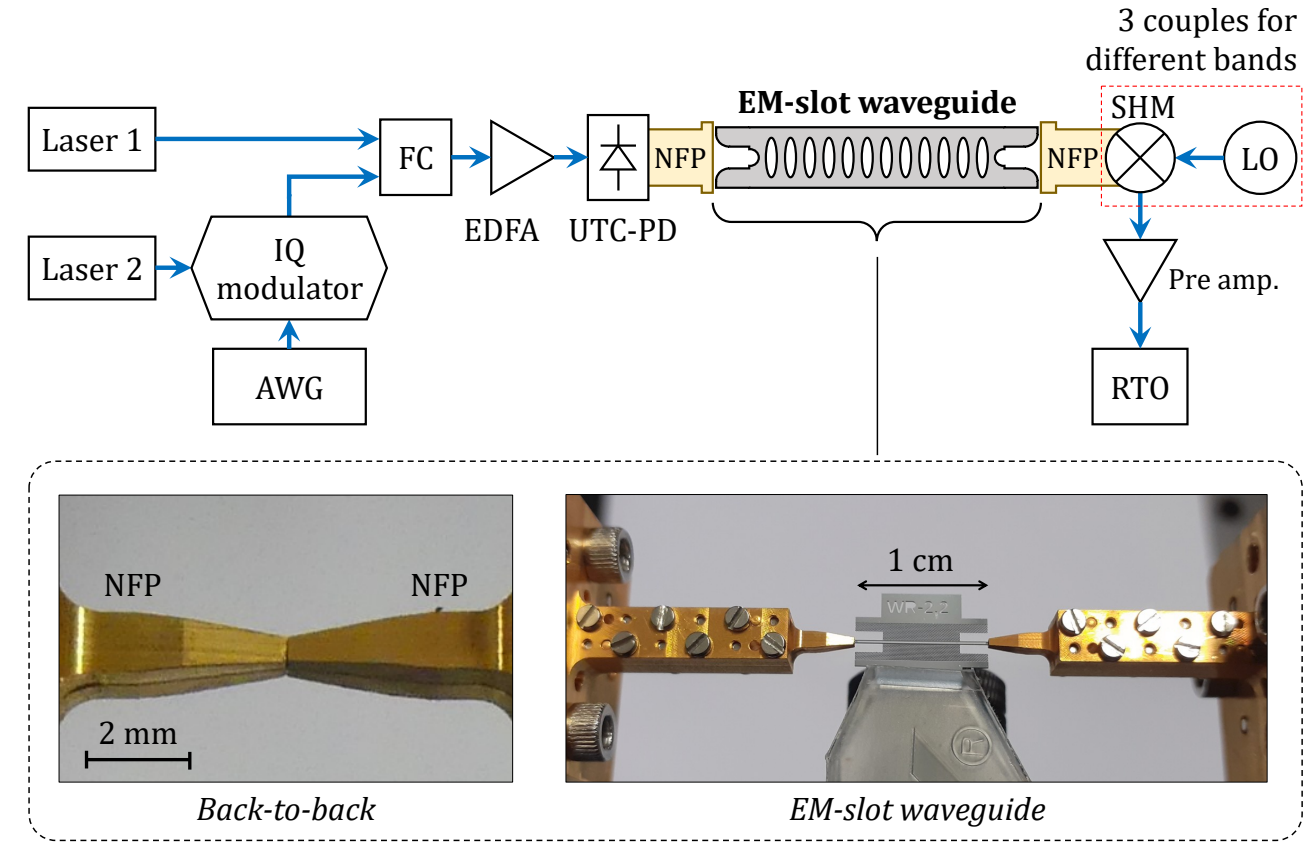

FIG. 12. Block diagram of communications with I/Q modulation. Near-field probes for WR-2.2 band are fixed for all carrier frequencies. On the receiver side, three sets of SHM/LO couples are used to cover multiple channels.

40 GHz, falls within the detection bandwidth of a 70-GHz real-time oscilloscope (UXR, Keysight). Signal demodulation is performed using Keysight's Vector Signal Analysis (VSA) software.[59] Digital processing in the VSA includes equalization through digital filtering to compensate for the system's frequency-dependent response, particularly the roll-off effect of the SHM.

### B. I/Q modulated communications with multiple channels

The frequency range is determined by the broadband response of the channel waveguide, spanning from 330 GHz carrier up to 600 GHz, as shown in the Fig. 13(a). In order to fully exploit this wide spectrum for communications, 14 channels are allocated at various carrier frequencies, as listed in Table I. While the lasers are tuned to change the carrier frequency emitted by the photonic-based transmitter, a set of three receivers is used to cover the broad terahertz bandwidth. The first receiver is a WR-2.2 subharmonic mixer (SHM), enabling the detection of channels 1~8 (330 to 470 GHz). The second receiver, used for channels 9~11, covers 480 to 550 GHz (carriers 490/515/540 GHz) is composed of SHM integrated in a WR-1.9 (400–600 GHz) waveguide. The last receiver is a WR-1.5 SHM, to cover the last 3 channels (12~14) corresponding to the 570/585/600 GHz carriers.

Each carrier frequency is tested in two configurations. First, the "back-to-back" (B2B) case, where the near-field probes of the transmitter and receiver are in direct contact, sets the reference performance in terms of the error-vector magnitude (EVM) metric.[60] Second, the EM slot waveguide is employed in the system and the EVM performance is measured again. Several quadrature amplitude modulation (QAM) formats are used to assess the performance of the channel waveguide, consisting of 16-QAM, 32-QAM, and 64-QAM. Figure 13(b) shows the maximum data rate that the channel waveguide can support at each carrier frequency. The achievable data rate ranges from 32 to 80 Gbit/s, resulting in the aggregated throughput of 809 Gbit/s, enabling the potential of the waveguide for ultra-high-speed terahertz interconnects. Noticeably, the data rates measured with the WR-1.5 SHM are the lowest, likely due to the ripples observed in the transmittance and group delay around 600 GHz, as shown in Fig. 11. Despite the degradation, optimum bit-error rates (BER) remain below the hard-decision forward-error-correction (HD-FEC) limit (BER $< 3.8 \times 10^{-3}$), as shown in Fig. 13(c). For 16-QAM modulation, both configurations achieve BER values well below the HD-FEC threshold. For higher-order modulations (32-QAM and 64-QAM), a slight degradation is observed even in the reference measurement, primarily attributed to the limitations of the experimental testbed rather than the intrinsic response of the channel waveguide. Figures 13(d-f) show the EVM performance versus relative power at the transmitter of the channel waveguide and the B2B at three carrier frequencies (370/490/600 GHz). The relative power is defined as the power variation with respect to the maximum power used for each carrier, and the power penalty is extracted as the relative power variation required to achieve the same EVM as the B2B case. The measured power penalty ranges from 1 to 4 dB, which is consistent with the transmittance characteristics of the waveguide. In all cases, the channel waveguide exhibits good performance with a limited EVM degradation, as shown in Table I.

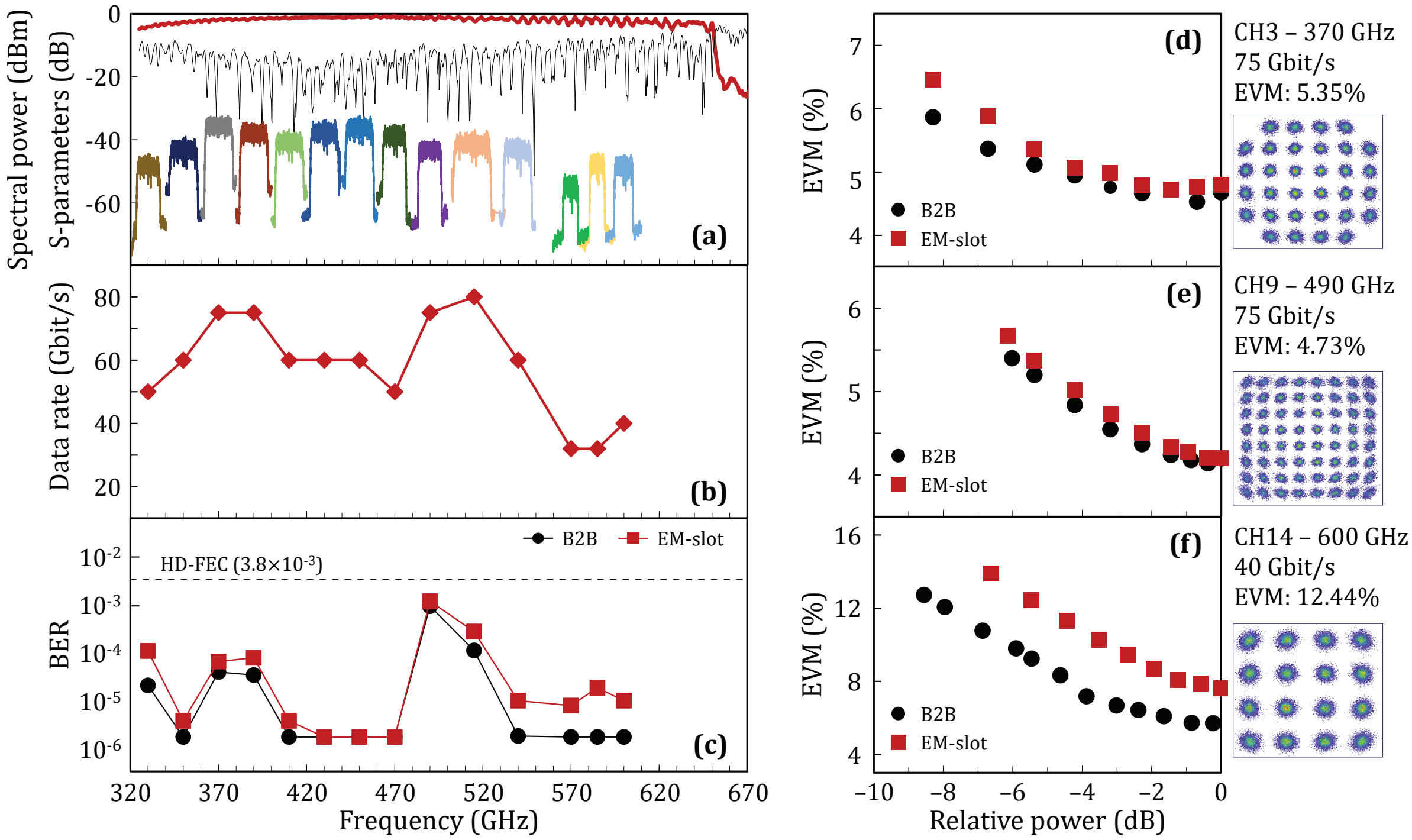

FIG. 13. Communications performance. (a) Measured absolute received power spectra at 14 allocated frequencies. S-parameters are also embedded. (b) Data rate of 14 channels, with aggregated data rate of 809 Gbit/s. Maximum/minimum: 80/32 Gbit/s. (c) BER at each carrier frequency for B2B and with EM-slot waveguide. EVM versus relative transmitted power of B2B (black circles) and with EM-slot waveguide (red squares) at (d) 370 GHz (channel 3, 32-QAM), (e) 490 GHz (channel 9, 64-QAM), and (f) 600 GHz (channel 14, 16-QAM). Constellation diagram at corresponding EVM are also included. Relative power: power variation with respect to the maximum power used for each carrier.

TABLE I. Frequency allocation, modulation types, and data rates of 14 channels. Aggregated data rate: 809 Gbit/s.

| SHM's band | WR-2.2 | | | | | | | | WR-1.9 | | | WR-1.5 | | |
|---|---|---|---|---|---|---|---|---|---|---|---|---|---|---|
| Channel | 1 | 2 | 3 | 4 | 5 | 6 | 7 | 8 | 9 | 10 | 11 | 12 | 13 | 14 |
| $f_c$ (GHz) | 330 | 350 | 370 | 390 | 410 | 430 | 450 | 470 | 490 | 515 | 540 | 570 | 585 | 600 |
| Mod. type (-QAM) | 16 | 16 | 32 | 32 | 16 | 16 | 16 | 16 | 64 | 16 | 16 | 16 | 16 | 16 |
| Power penalty (dB) | 3.5 | 3.1 | 1.3 | 2.2 | 1.9 | 1.6 | 1.0 | 1.9 | 1.0 | 1.4 | 2.2 | 4.1 | 3.9 | 3.1 |
| Data rate (Gbit/s) | 50 | 60 | 75 | 75 | 60 | 60 | 60 | 50 | 75 | 80 | 60 | 32 | 32 | 40 |
| EVM (%) | 8.5 | 7.5 | 5.1 | 5.0 | 7.3 | 6.8 | 6.2 | 7.0 | 4.2 | 9.3 | 7.4 | 7.4 | 8.0 | 7.6 |

Table II compares the high-speed communication performance of state-of-the-art all-dielectric terahertz interconnects from 0.1 to 1 THz. A valley photonic-crystal waveguide demonstrates a data rate of 108 Gbit/s, yet their bandwidth is only about 30 GHz (~10% fractional bandwidth) owing

TABLE II. State-of-the-art all-dielectric devices for terahertz communications, sorted by data rates.

| Interconnect | Frequency | Tx | Rx | Distance | Modulation | No. of channels | Data rate (Gbit/s) | | BER |
|---|---|---|---|---|---|---|---|---|---|
| | (GHz) | | | (mm) | | | Max* | Total | |
| ***Single channel*** | | | | | | | | | |
| EM WG[28] | 335 | UTC-PD | SBD | 30 | OOK | 1 | 28 | 28 | $9 \times 10^{-12}$ |
| PC WG[61] | 350 | UTC-PD | RTD | 6 | OOK | 1 | 32 | 32 | $2 \times 10^{-12}$ |
| Taperless WG[51] | 350 | UTC-PD | RTD | 20 | OOK | 1 | 32 | 32 | $9 \times 10^{-12}$ |
| EM WG (on-chip)[62] | 315 | UTC-PD | RTD | 50 | 32-QAM | 1 | 100 | 100 | $3.6 \times 10^{-3}$ |
| VPC WG[63] | 317 | UTC-PD | SHM | 10 | 16-QAM | 1 | 108 | 108 | $7.4 \times 10^{-4}$ |
| ***Multiple channels*** | | | | | | | | | |
| Multi-mode WG[64] | 165 | CMOS | CMOS | 52 | OOK | 3 | 24 | 65 | $10^{-12}$ |
| Diplexer VPC WG[65] | 305<br>321.6 | UTC-PD | SHM | 38 | 32-QAM | 2 | 75 | 150 | $5 \times 10^{-3}$ |
| Polarization multiplexer[35] | 310 | UTC-PD | FMBD | 20 | 16-QAM | 2 | 80<br>100 | 155<br>190 | $< 3.6 \times 10^{-3}$<br>$< 2.0 \times 10^{-2}$ |
| Hollow WG[13] | 500–724 | UTC-PD | SHM | B2B | 16,32,64-QAM | 12<br>14 | 100<br>120 | 887<br>1,041 | $< 8.0 \times 10^{-4}$<br>$< 8.3 \times 10^{-3}$ |
| **This work** | **330–600** | **UTC-PD** | **SHM** | **10** | **16,32,64-QAM** | **14** | **80** | **809** | $< \mathbf{1.2 \times 10^{-3}}$ |

*Maximum data rate of a single channel.
WG: waveguide; PC: photonic crystal; VPC: valley photonic crystal;
SBD: Schottky barrier diode; RTD: resonant tunneling diode; FMBD: Fermi-level managed barrier diode; SHM: sub-harmonic mixer.

to intrinsic photonic-bandgap dispersion.[63] In contrast, EM-based waveguides support much broader bandwidths (approximately 40%), and deliver single-link data rates up to 100 Gbit/s. Most demonstrations, however, remain below 350 GHz, limited by the availability of efficient Tx at higher frequencies. With the advent of the 600-GHz UTC-PD source and SHM detectors, B2B transmission through WR-1.5 hollow waveguides has surpassed 1 Tbit/s.[13] In advanced photonic integrated circuits, ultra-broadband and low-loss interconnects are essential to facilitate high data rates with high energy efficiency. Our proposed EM-slot waveguide supports operation from 330 GHz to 600 GHz with an average loss of –1.7 dB. Building on these advances, the proposed platform fully exploits its ultra-broadband characteristics to accommodate 14 channels with BER below the HD-FEC threshold, enabling aggregate data rates of 809 Gbit/s from 330 GHz to 600 GHz, thereby approaching the testing system capability with a power-penalty in range of 1–4 dB. As discussed later in Section V-B, the intrinsic bandwidth of the waveguide extends up to 700 GHz, covering most of the entire WR-2.2 and WR-1.5 bands. Further optimization of the waveguide geometry in the extended frequency band is expected to reduce waveguide dispersion and reflection losses, targeting terabit-per-second terahertz communication.

## V. DISCUSSIONS ON EFFECTIVE-MEDIUM-SLOT WAVEGUIDES

### A. Comparison between dielectric waveguides

Table III compares performances of the proposed substrateless slot waveguide with previously reported terahertz dielectric waveguides coupled to metallic hollow waveguides. Waveguides with supporting substrates, typically based on silicon-on-insulator platforms, exhibit relatively high attenuation (>1.4 dB/cm), mainly due to substrate-induced absorption and radiation leakage.[47,66] In contrast, substrateless waveguides fabricated from high-resistivity silicon demonstrate much lower propagation loss, with average values on the order of 0.1 dB/cm. Conventional dielectric waveguides require tapered transitions to achieve mode and impedance matching to metallic hollow waveguides.[28,37,67] In the WR-3.4 band (220–330 GHz), the multimode interferometer waveguide reduces the insertion length from a conventional 3 mm to 0.26 mm by employing a trapezoidal slot coupler.[48] This coupler operates based on a quarter-wave matching principle, where reflections at the two slot interfaces destructively interfere. However, because the guided wavelengths at low frequencies exceed the interference range, the coupler must be inserted into the hollow core to suppress radiation leakage. To achieve an insertion-free coupling mechanism, we designed taperless couplers using apodized-slot geometries to broaden the matching bandwidth to 36%.[51] Nevertheless, this structure still relies on subsequent tapered waveguides, leading to additional loss due to multiple mode-conversion stages. In contrast, the presented slot waveguide employs an EM channel for broadband modal matching, directly converting the mode from the slot coupler to the guiding channel without relying on auxiliary leakage-suppression structures, achieving a 40% fractional bandwidth.

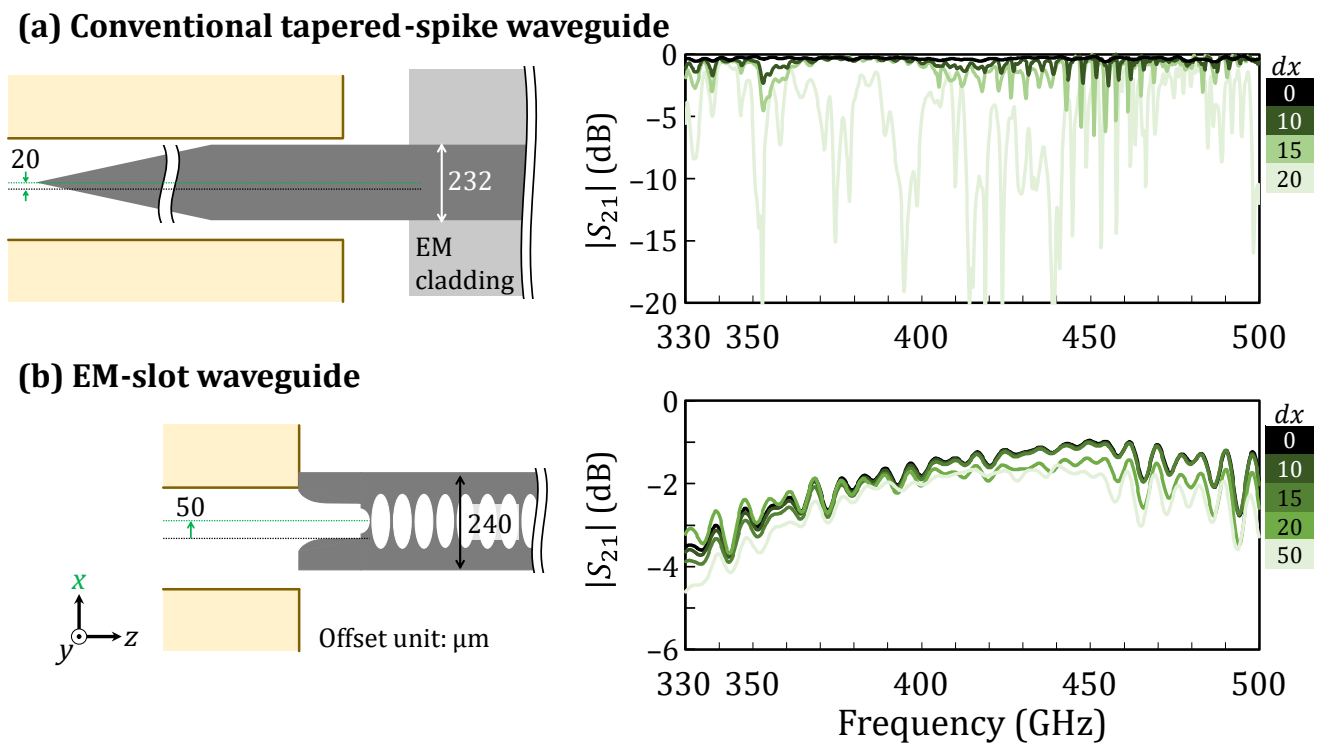

FIG. 14. Simulation models and transmission coefficients for $x$-axis misalignment offsets of (a) conventional tapered-spike waveguide and (b) EM-slot waveguide. The offset unit is in μm.

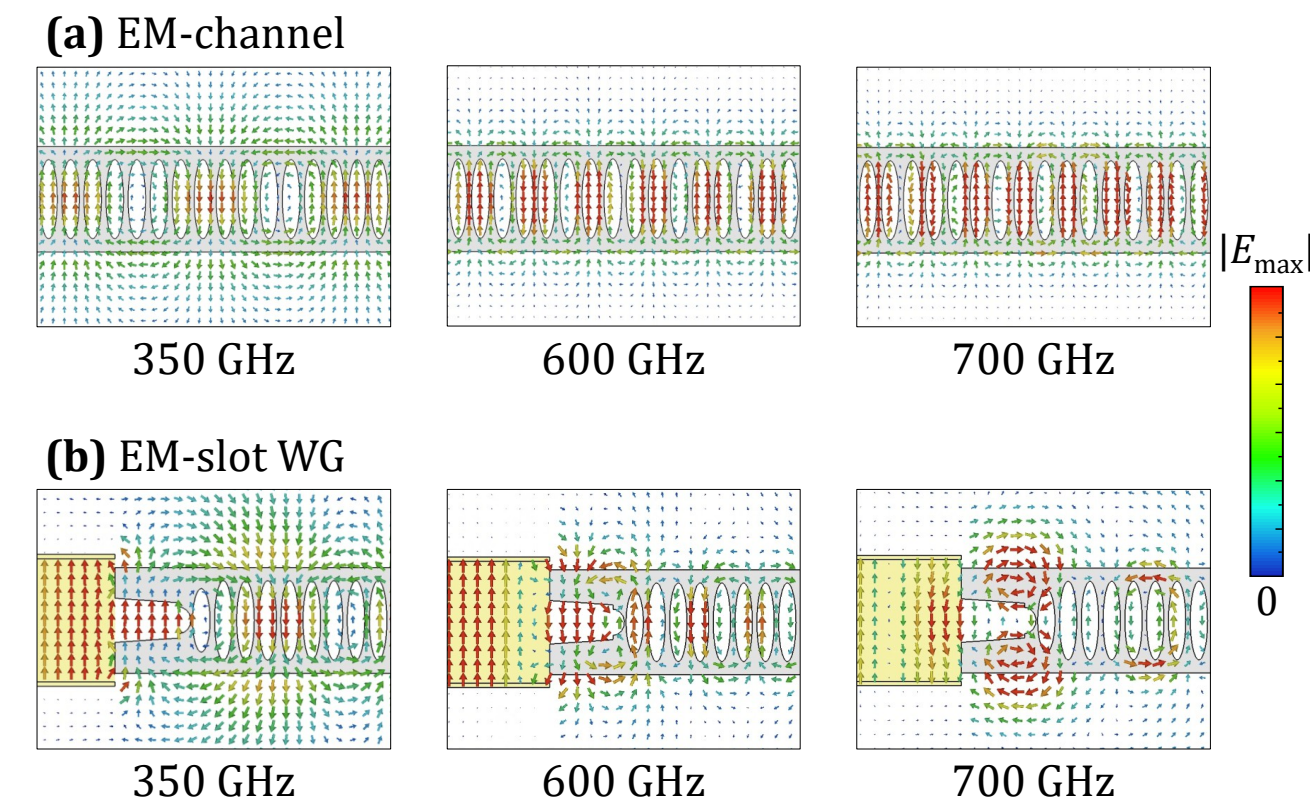

FIG. 15. Observations of higher-order modes. Instantaneous vector $E$-field distributions of (a) EM-channel fed by waveguide ports and (b) EM-slot waveguide (without EM cladding) coupled to metallic hollow waveguides.

It is worth noticed that conventional tapered-spike couplers demand alignment along all three Cartesian axes ($x$, $y$, and $z$). In contrast, our EM-slot waveguide geometry requires only $x$-$y$ alignment, i.e., the alignment at the interface of the output of the hollow waveguide. Particularly, minor misalignment along the $x$-axis is no longer affected by silicon-metal interactions that would generate strong standing waves and dramatically degrade transmittance, as shown in Fig. 14. The sensitivity to such misalignment in three axes are further discussed in *Supplementary Material–S2*. The EM-slot waveguide with a taper-free coupler therefore provides improved robustness for practical implementation in integrated devices, particularly for large-scale production. In addition, avoiding insertion of a taper into the hollow core reduces the risk of damaging the waveguide surface, which could introduce surface roughness and significantly reduce the transmittance, as demonstrated in *Supplementary Material–S2*.

Furthermore, the array of air holes can be transformed into a homogeneous silicon channel (no air holes) by gradually tapering the sizes of air holes (as described in *Supplementary Material–S3*), making it compatible with silicon-integrated platforms[33,35,36,68], while providing a robust interface with external hollow waveguide-based systems. We can also integrated compact electronic devices,[69,70] to realize advanced active components, in which detailed coupling models can be found in *Supplementary Material–S4*. Finally, packaging of this channel waveguide does not require metallic-hollow-waveguide interface,[71] enabling lower fabrication complexity and better scalability with 3D-printing technologies.[72]

### B. Multi-mode at high-frequency operation

To evaluate mode distributions beyond WR-2.2 frequency band, we first analyze the intrinsic mode properties of the elliptical air-hole channel without the slot couplers. The simulation was done by using the full wave simulator, in which the details are shown in *Supplementary Material–S5*. The coupling efficiencies of higher-order modes are less than 10%. This shows that the fundamental mode dominates over the entire WR-2.2 and WR-1.5 bands, in which the $E$-field distribution at different frequencies are shown in Fig. 15(a). After introducing the slot coupler, the waveguide is fed by a metallic hollow waveguide with the fundamental TE mode. As illustrated in Fig. 15(b), the excited field consistently corresponds to the fundamental slot mode within the 330–600 GHz communication band. The electric field is strongly confined within the slot region and then gradually evolves into the EM-channel mode, indicating smooth modal transformation across the interface. As the frequency increases beyond approximately 650 GHz, transmission degradation becomes noticeable. For example, the field at 700 GHz couples to the silicon beams of the slot coupler rather than confined inside the slot. This behavior is associated with the excitation of higher-order modes. In particular, the slot coupler was originally optimized for operation up to 500 GHz, which limits its high-frequency performance. This leads to excitation of higher-order modes and the decreased transmittance beyond the communication bandwidth, as shown by degradation over 690 GHz in Fig. 16(a). To evaluate the excited mode of the EM-slot waveguide coupling to metallic hollow waveguides, we calculate the overlap integral between the mode of the EM-slot waveguide and the isolated EM-channel as follows:

$$\eta = \frac{\iint E_{\text{EM-slot}} \cdot E^{*}_{\text{ref}}\, \mathrm{d}A}{\iint |E_{\text{EM-slot}}|^{2}\, \mathrm{d}A \cdot \iint |E_{\text{ref}}|^{2}\, \mathrm{d}A}, \tag{4}$$

where $E_{\text{EM-slot}}$ and $E_{\text{ref}}$ are the $E$-fields of the EM-slot waveguide and isolated EM-channel, respectively. As shown in Table IV, the calculated overlap integrals $\eta$ from 330–650 GHz is above 95%, demonstrating the dominant slot mode are excited in the waveguide. However, the overlap integral at 700 GHz significantly reduces 66.3%, showing that the slot mode gradually vanished, which is likely to the mode mismatch of the slot coupler.

TABLE III. Comparison of terahertz silicon waveguides coupled to metallic hollow waveguides, sorted by insertion length.

| Waveguide type | Frequency (GHz) | 3-dB bandwidth (%) | Propagation loss (Average/Max) (dB/cm) | $\lvert S_{21}\rvert_{\max}$ (dB) | Insertion length (mm) |
|---|---|---|---|---|---|
| ***With supporting substrate*** | | | | | |
| SIIG WG[66] | 215–335 | 44 | 1.43 / - | –2 | 2.72 |
| Slot WG on a SBQ platform[47] | 500–570 | 13 | 1.68 / 4 | –2.9 | 0.75 |
| MEMS slot WG[44] | 224–326 | 37 | 1.5 / - | –11 | 0 |
| ***Substrateless*** | | | | | |
| Photonic crystal WG[67] | 322–350 | 8 | <0.05 / 0.1 | ~0 | 6 |
| EM-cladded WG[28] | 260–390 | 42 | 0.05 / 0.09 | –0.1 | 3 |
| E-skid WG[37] | 220–330 | 40 | 0.1 / 0.37 | –0.05 | 2.7 |
| MMI WG[48] | 237–325 | 31.3 | - | –0.3 | 0.26 |
| Taperless-interfaced WG[51] | 349–500 | 36 | - | –1.0 | 0 |
| **This work** | **335–500** | **40** | **0.05 / 0.38** | **–0.9** | **0** |

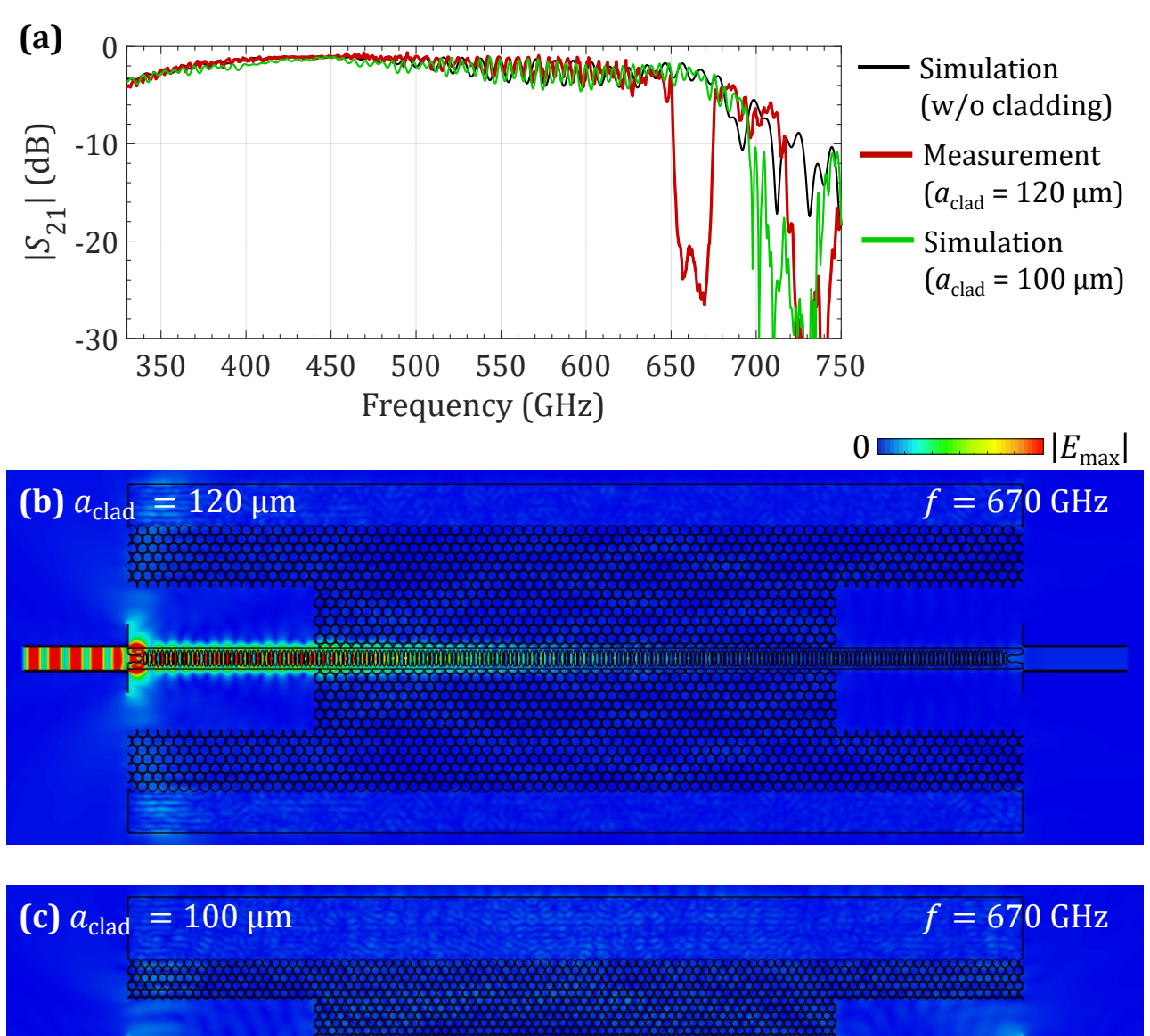

FIG. 16. (a) Transmittance of EM-slot waveguides. Lattice constant of EM cladding $a_{\mathrm{clad}}$ is reduced to 100 μm for improved structure (shown in green). Maximum simulated $E$-field intensities of EM-slot waveguides with (b) $a_{\mathrm{clad}} = 120$ μm and (c) $a_{\mathrm{clad}} = 100$ μm.

In addition, the measured transmission exhibits a dip at around 650–680 GHz, as shown in Fig. 16(a). The corresponding $E$-field intensity distribution at 670 GHz is illustrated in Fig. 16(b), indicating that the mode cannot propagate when the wave reaches the EM cladding region. This behavior suggests that the EM cladding with a lattice constant of $a_{\mathrm{clad}} = 120$ μm no longer behaves as an effective medium at higher frequencies. Instead, the periodic air-hole lattice begins to exhibit photonic-crystal characteristics, leading to the formation of a photonic band edge with a reduced group velocity and a mode gap.[27,73] This issue can be mitigated by reducing the lattice constant of the EM cladding to $a_{\mathrm{clad}} = 100$ μm. In this case, the periodic structure remains in the EM regime over the frequency band of interest, as evidenced by the wider transmission bandwidth and the continuous wave propagation along the waveguide in Figs. 16(a) and (c). However, the transmittance is significantly reduced at frequency beyond 700 GHz. This is mainly because of the effective-medium approximation applied to the EM channel is frequency dependent, scaling up the operation frequency requires the reduction of the air-hole period. The minimum distance is estimated to be less than a quarter of the guided wavelength at the highest frequency (750 GHz) in the EM channel, i.e., the period $d$ is approximated at $d \leq c/(4 f_{\mathrm{high}} n_{\mathrm{EM}})$ (= 44 μm). The current period of the air-hole array is 50 μm, which is slightly higher than the estimated value.

TABLE IV. Overlap integral between the mode of EM-slot waveguide and that of isolated EM-channel.

| Frequency (GHz) | 350 | 400 | 450 | 500 | 550 | 600 | 650 | 700 | 750 |
|---|---|---|---|---|---|---|---|---|---|
| $\eta$ (%) | 98.5 | 98.9 | 98.7 | 98.0 | 98.8 | 98.6 | 95.5 | 66.3 | 71.3 |

## VI. CONCLUSION

We have proposed a substrateless slot waveguide along with an effective coupler to interface with rectangular waveguides. A transition of a slot coupler is modified for refractive index and mode profile matching, resulting in $E$-field confinement within the air gap. An EM slot formed by an elliptical air-hole array is introduced for connecting two beams of the slot coupler and emulating an air-like refractive index. This platform offers broadband and low-loss characteristics, enabling high-capacity terahertz communications. The waveguide is experimentally validated in the WR-2.2 band, exhibiting an average propagation loss of 0.05 dB/cm and a maximum coupling efficiency of 90%, and maintains good matching up to 600 GHz. Leveraging this wide operational bandwidth, data transmission through multiple channels, spanning 330–600 GHz, is demonstrated using a broadband UTC-PD as a transmitter with three SHM receivers for different frequency bands. By using QAM, the channel waveguide exhibits high data rate with limited EVM degradation, achieving an aggregate data rate of 0.8 Tbit/s with the BER below the HD-FEC limit for all channels. The interface is also robust to misalignment compared with a conventional tapered spike, yielding an efficient approach for packaging dielectric waveguide. Furthermore, the elliptical air holes can be progressively tapered,[50] transforming the EM channel into various passive components. This versatility enables efficient interconnection for ultra-high-speed communications and integrated sensing, as conceptualized in Fig. 1, paving the way for applications toward 6G and beyond.

## SUPPLEMENTARY MATERIAL

See supplementary material for more details on (1) design optimization of effective-medium slot waveguide, (2) alignment test, (3) utilization of THz substrateless slot waveguide for interconnects, (4) integration with electronic components, and (5) higher-order mode analysis.

## ACKNOWLEDGMENTS

This work was supported in part by the JST Core Research for Evolutional Science and Technology (JPMJCR21C4), the Grants-in-Aid for Scientific Research (24H00031), and the Australia-Japan Foundation, Australia (AJF2021044). G.D. and P.S. state that the characterization testbeds are supported by France 2030 programs, PEPR (Programmes et Equipements Prioritaires pour la Recherche) and CPER Wavetech. The PEPR is operated by the Agence Nationale de la Recherche (ANR), under the grants ANR-22-PEEL-0006 (FUNTERA, PEPR 'Electronics') and ANR-22-PEFT-0006 (NF-SYSTERA, PEPR 5G and beyond - Future Networks). The ANR project PICOTE (ANR-23-CE24-0023) also supported the development of the 600 GHz testbed. The Contrat de Plan Etat-Region (CPER) WaveTech is supported by the Ministry of Higher Education and Research, the Hauts-de-France Regional council, the Lille European Metropolis (MEL), the Institute of Physics of the French National Centre for Scientific Research (CNRS) and the European Regional Development Fund (ERDF). The authors acknowledge support from CDP C2EMPI, the French State under the France-2030 program, the University of Lille, the Initiative of Excellence of the University of Lille, the European Metropolis of Lille for their funding and support of the R-CDP-24-004-C2EMPI project.

# *Supplementary information*
# Terahertz Communications Using Effective-Medium-Slot Waveguides

Nguyen H. Ngo[1,*], Weijie Gao[1], Guillaume Ducournau[2], Hadjer Nihel Khelil[2], Rita Younes[2], Pascal Szriftgiser[3], Hidemasa Yamane[4], Yoshiharu Yamada[4], Shuichi Murakami[4], Withawat Withayachumnankul[5], Masayuki Fujita[1]

[1] Graduate School of Engineering Science, The University of Osaka, Toyonaka 560-8531, Japan
[2] CNRS, UMR 8520 – IEMN, Institut d'Electronique de Microélectronique et de Nanotechnologie, Université de Lille, Lille 59000, France
[3] Laboratoire de Physique des Lasers, Atomes et Molécules, UMR CNRS 8523 PhLAM, Université de Lille, Lille 59000, France
[4] Osaka Research Institute of Industrial Science and Technology, Izumi 594-1157, Japan
[5] Terahertz Engineering Laboratory, Adelaide University, Adelaide SA 5005, Australia
[*]*ngo.h-nguyen.es@osaka-u.ac.jp (N.N.), gao.weijie.es@osaka-u.ac.jp (W.G.), fujita@ee.es.osaka-u.ac.jp (M.F.)*

## 1. Design optimization of effective-medium slot waveguide

The effective-medium (EM)-slot waveguide consists of two parts: an EM-channel and slot couplers at the two ends. Initial design parameters of the EM-slot waveguides are visualized and listed in Fig. S1.

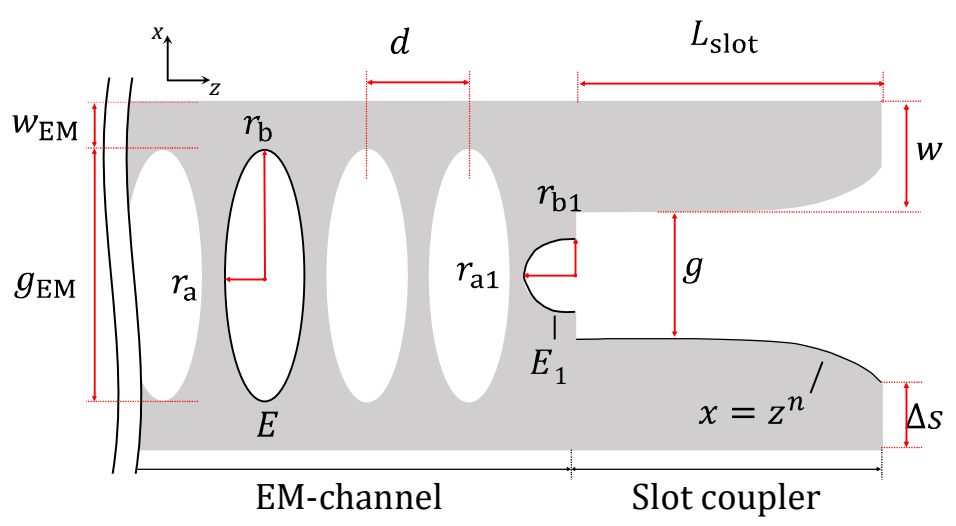

| *EM channel* | | *Slot coupler* | |
|---|---|---|---|
| Channel width | $w_{EM}$ | Beam width | w |
| Channel gap | $g_{EM}$ (= $2r_b$) | Length | $L_{slot}$ |
| Total width | $W_{total}$ | Gap | $g$ |
| Period | $d$ | Curvature | $n$ |
| Channel ellipse | | Offset | Δs |
| Minor radius | $r_a$ | | |
| Major radius | $r_b$ | | |
| First ellipse | | | |
| Minor radius | $r_{a1}$ | | |
| Major radius | $r_{b1}$ | | |

***Fig. S1*** *Schematic with design parameters of EM-slot waveguide.*
ALT TEXT: Schematic of the EM-slot waveguide. Design parameters for EM channel and slot coupler are listed.

The thickness of the waveguide ($h$) is fixed at 200 μm, which is the thickness of used silicon wafer. The total width ($W_{total}$) of the waveguide is equal to $2w+g$ (= $2w_{EM}+g_{EM}$). The merit for optimization of the parameters is to maximize the average transmittance ($S_{21}$) and minimize the average reflectance ($S_{11}$) in the WR-2.2 band (330–500 GHz).

Since there are several parameters, the EM-channel with a homogeneous permittivity is determined first. The initial conditions for the channel width are fixed at 260 μm, which is close to the width of the core of the metallic hollow waveguide (279 μm). The period of elliptical air holes ($d$) is set to 50 μm, which is much shorter than quarter of the shortest guided wavelength (i.e., 500 GHz). The gap between two elliptical air holes should be larger than 10 μm, which is limited by fabrication to avoid over-etching. This gives the minor radius ($r_a$) to be 20 μm with estimated relative permittivity of the EM-channel is ($\varepsilon_x$, $\varepsilon_y$, $\varepsilon_z$) = (3.65, 5.04, 3.65) [S1]. The gap of the EM-channel ($g_{EM}$) is varied, and the transmittance is shown in Fig. S2. In order to avoid under-etching of the gap, the minimum $g_{EM}$ is set to 40 μm. As seen from the graph, the highest average transmittance is obtained when the gap is 140 μm. As a result, initial value of channel width ($w_{EM}$) and gap ($g_{EM}$) is set to 60 μm and 140 μm, respectively.

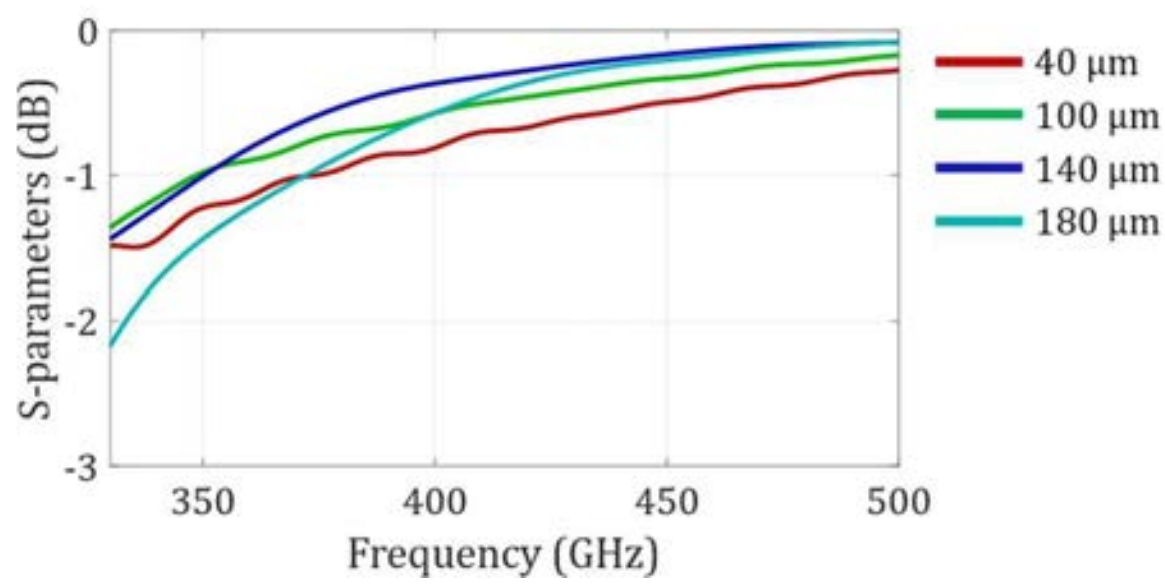

***Fig. S2*** *S-parameters of the EM-slot channel with various channel gaps* $g_{EM}$.

ALT TEXT: The transmittance plot shows increasing transmission with frequency and improved performance for an optimized slot gap.

**Optimization of slot couplers**

After the EM-channel is validated in the previous step, metallic hollow waveguides (HWG) are employed to feed the EM-channel. Since there is a large refractive index mismatch between the hollow core and the EM-channel, a slot coupler is employed to enhance coupling efficiency and reduce the reflection. In order to optimize parameters of the waveguide, the CST built-in covariance matrix adaptation evolution strategy (CMA-ES) optimizer is employed. The optimization steps, from step 1 to step 4, are listed below. Since there are a number of parameters, we follow from step 1 to step 3 to identify initial design parameters, as well as the design constraints. After that, the CMA-ES optimizer is applied for the final design in step 4.

***1. Optimization of slot coupler connecting to EM channel***

The slot coupler consisting of two silicon trapezoids with a length ($L_{slot}$) separated by a gap ($g$). The cutting offset ($\Delta s$) is initially set to 50 μm. The metallic hollow waveguide couples to the slot coupler, as shown in Fig. S3. In this step, $L_{slot}$ and $g$ are optimized by evaluating average S-parameters in the whole WR-2.2 band, as shown in Fig. S3 (b-c). From the heatmaps, the transmittance peaks at the optimal slot length of 150 μm, whereas varying the slot gap from 60 to 140 μm results in slight differences in transmittance and reflection. The optimum S-parameters are achieved when the slot length and gap are 150 μm and 140 μm.

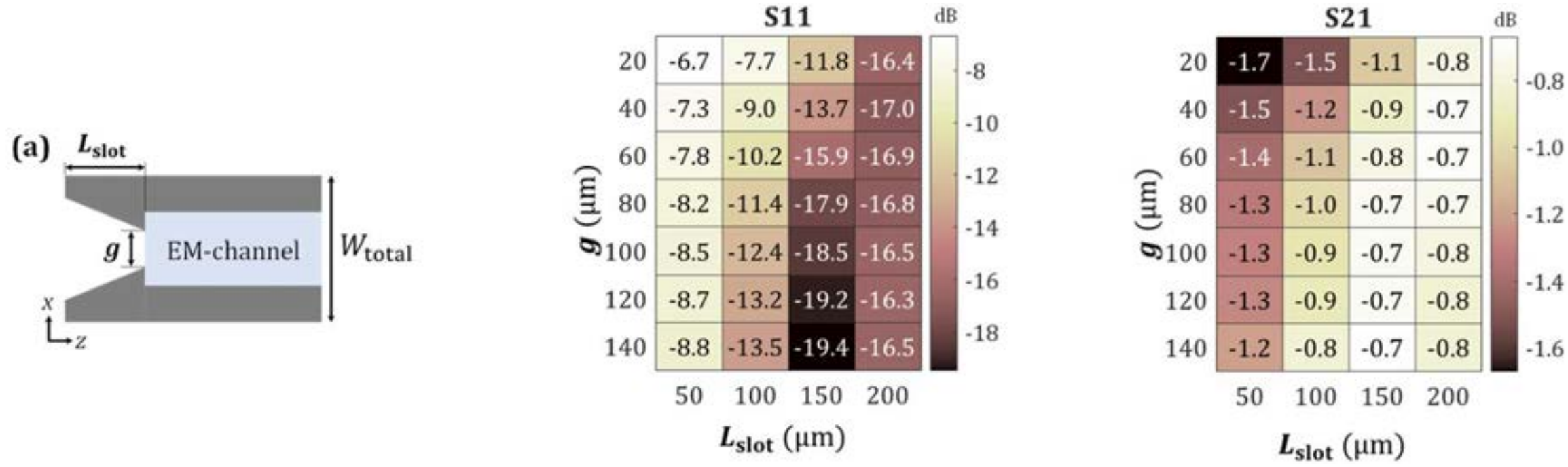

***Fig. S3*** *Optimization of the length and gap of the slot coupler. (a) Schematic of the slot couplers. (b) Average reflection coefficient and (b) average transmission coefficient of the system.*

ALT TEXT: Schematic of the slot couplers with two parameters: slot length and channel gap. Heatmaps of reflection and transmission coefficients show the variations, where minimum reflectance and maximum transmittance can be chosen for initial conditions.

***2. Optimization of curvature of slot coupler***

The homogeneous EM-channel is substituted by the array of elliptical air holes. To further optimize S-parameters, the curvature of the slot is determined by an $n^{th}$-order polynomial curve. The offset $\Delta s$ at the slot terminal is also adjusted. The schematic and S-parameters of this optimization are shown in Fig. S4. In this optimization, S-parameters reach their best when the polynomial order $n$ and offset $\Delta s$ are 7 and 50, respectively.

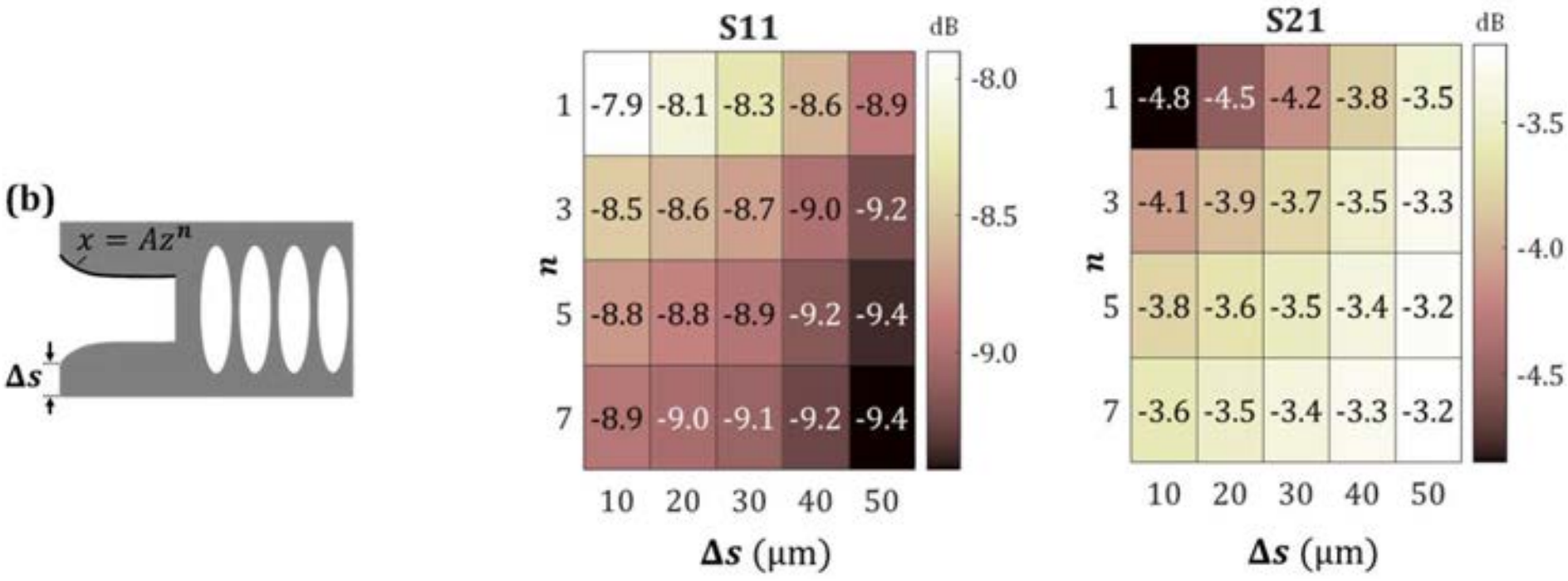

***Fig. S4*** *Optimization of the curvature and cutting offset of the slot coupler. (a) Schematic of the slot couplers. (b) Average reflection coefficient and (b) average transmission coefficient of the system.*

ALT TEXT: Schematic of the slot couplers with two parameters: slot curvature and slot offset. Heatmaps of reflection and transmission coefficients show the variations, where minimum reflectance and maximum transmittance can be chosen for initial conditions.

### 3. *Adjusting size of first elliptical hole*

To further smooth the transmittance, a semi elliptical air hole ($E_2$) is added to the edge of the slot coupler and EM channel, as shown in Fig. S5 (a). By varying the major radius $r_{b1}$, the transmittance becomes less fluctuating, as shown in Fig. S5 (b), which is likely due to an increased index and impedance matching, evidenced by lower reflectance.

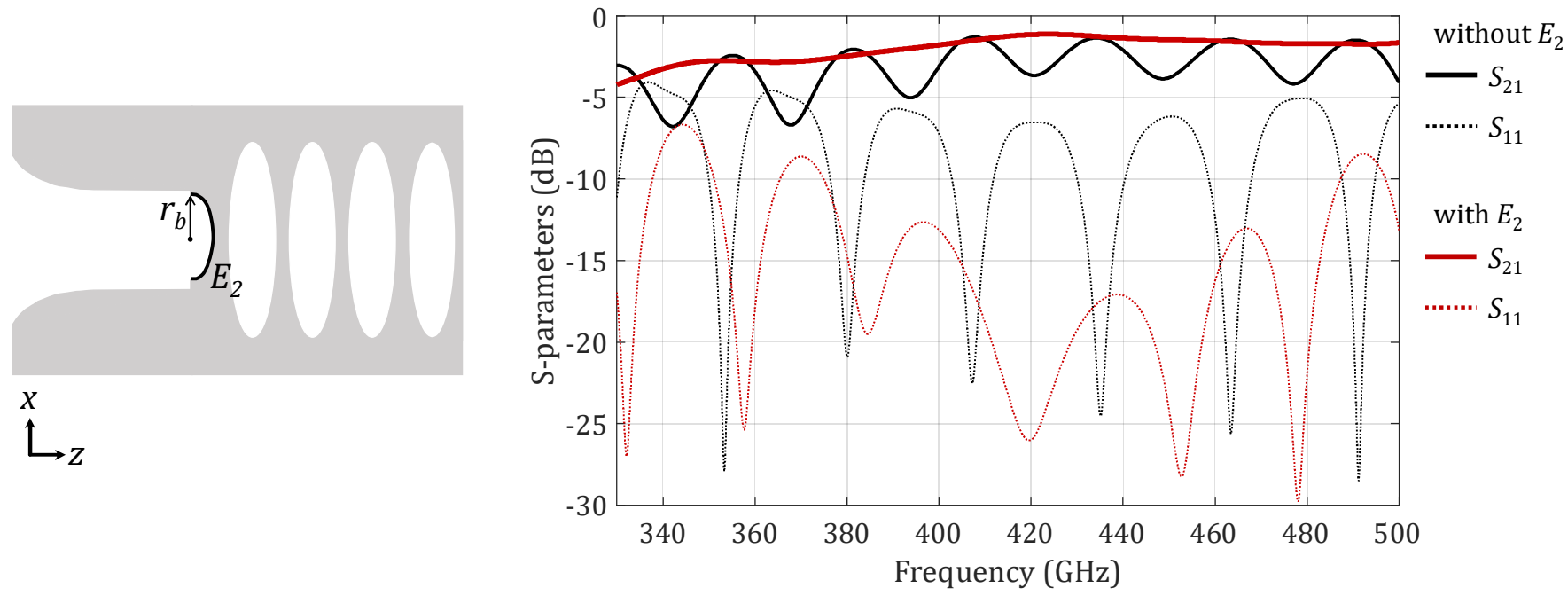

***Fig. S5*** *Optimization of the curvature and cutting offset of the slot coupler. (a) Schematic of the slot couplers. (b) Average reflection coefficient and (b) average transmission coefficient of the system.*

ALT TEXT: Schematic of the slot couplers, where the first elliptical air hole is modified. Line graphs of S-parameters show that the modified air hole helps smoothing the transmittance with reduced reflectance.

### 4. *Globally optimization of whole structure*

After obtaining the initial condition from step 3, the global optimization using built-in optimizer (CMA-ES) is utilized to achieve the highest average transmittance with lowest reflectance. In the final design, we also modified the second ellipse for better performance. *S*-parameters of the optimized EM-slot waveguide are shown in Fig. S6, showing that the final design has better performance. Figure S7 shows the schematic of the optimized waveguide. The values of design parameters before and after optimization are listed in Table S1.

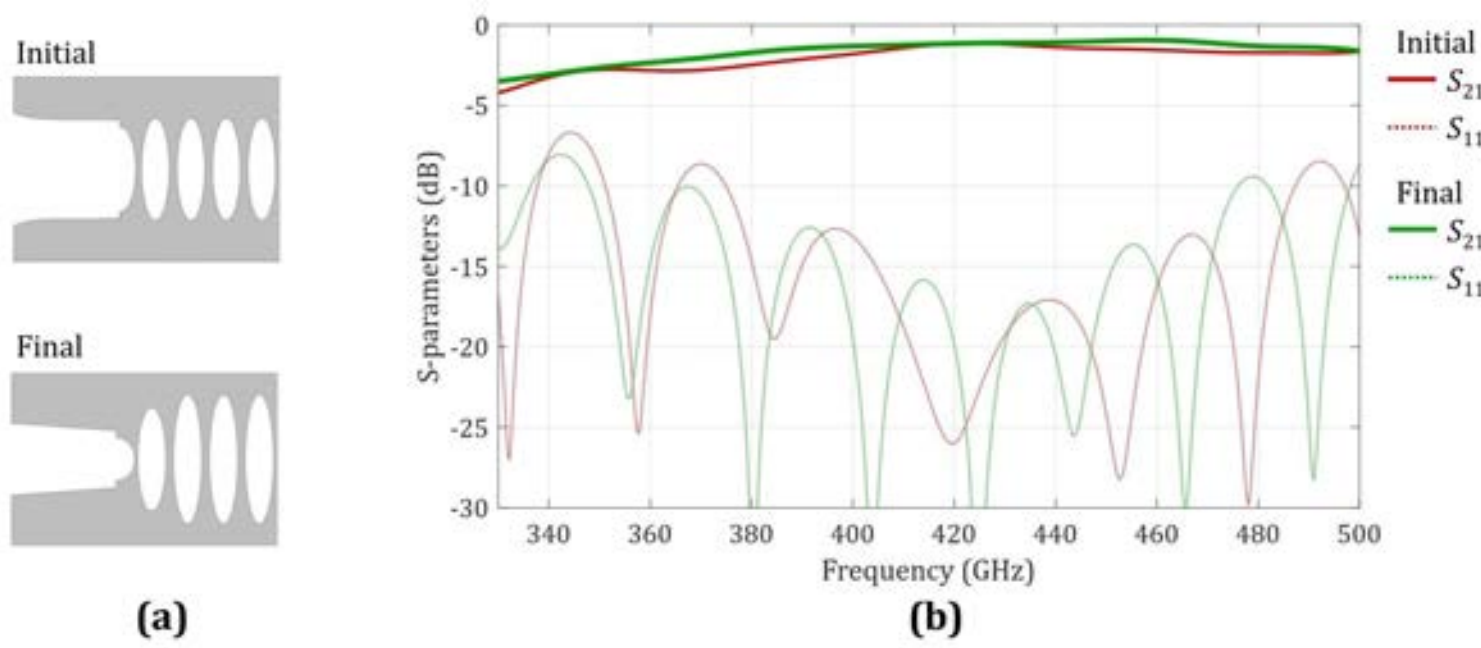

***Fig. S6*** *Optimization of the curvature and cutting offset of the slot coupler. (a) Schematic of the slot couplers. (b) Average reflection coefficient and (b) average transmission coefficient of the system.*
ALT TEXT: Schematic of the slot couplers before and after global optimization. Line graphs of S-parameters show that the final design has slight improvements than the initial design.

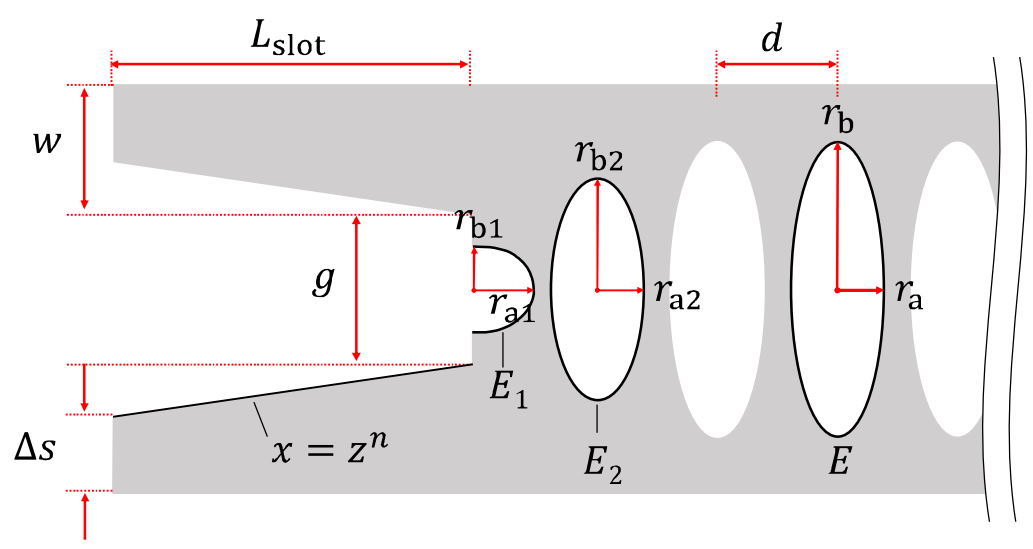

***Fig. S7*** *Schematic of the optimized EM-slot waveguide.*
ALT TEXT: Schematic of the slot couplers of the final design.

***Table S1*** *Initial and optimized design parameters of the EM-slot waveguide (unit: μm).*

| | $d$ | $W_{total}$ | $r_{a1}$ | $r_{b1}$ | $r_{a2}$ | $r_{b2}$ | $r_a$ | $r_b$ | $r_{b1}$ | $w$ | $L_{slot}$ | $g$ | $n$ | $\Delta s$ |
|---|---|---|---|---|---|---|---|---|---|---|---|---|---|---|
| *Initial* | 50 | 260 | 20 | 120 | 20 | 70 | 20 | 70 | 120 | 60 | 150 | 140 | 7 | 50 |
| *Optimized* | 50 | 240 | 26.4 | 60 | 19.8 | 72 | 19.8 | 90 | 60 | 80 | 145 | 80 | 1 | 70 |

ALT TEXT: Table data for all design parameters of the initial and final structure.

## 2. Alignment test

In conventional tapered spike designs, wave coupling occurs inside the core the metallic hollow waveguide to the silicon taper, achieving high coupling efficiency. Maintaining precise alignment of tapered spikes is practically challenging, as deviations of only tens of micrometers can significantly degrade coupling efficiency. For example, Ref. S2 reports simulated transmission coefficients ($|S_{21}|$) of 80–90%, whereas measured values after packaging drop to approximately 50%. Achieving perfect alignment typically requires meticulous manual adjustment under a microscope. In contrast, coupling to the proposed substrateless slot waveguide, which is completely outside the hollow core, results in partial energy leakage at the interface, slightly reducing transmittance. However, $|S_{21}|$ in this configuration is significantly less sensitive to misalignment than in the tapered-spike design. To elaborate the robustness of the substrateless slot waveguide for practical implementation, we compare the transmittances performance of the two structures through simulations.

***Model explanation***

Simulation models for the EM-clad waveguide and the EM-slot waveguide are shown in Fig. S7(a,b). Misalignment offsets along the *z*-, *x*-, and *y*-axes are represented in the top, middle, and bottom rows, respectively. The TE mode ($\boldsymbol{E}^{x}_{11}$) was excited at a WR-2.2 hollow waveguide port (559 μm × 279 μm). The taper length of the EM waveguide was 25 mm, and the EM-slot waveguide dimensions are provided in Fig. 5 of the main text. For the *z*-axis (propagation direction), the distance between the hollow-waveguide interface and the tapered spike edge was set to 300 μm to avoid strong reflections caused by metal proximity to the cladding. For the EM-slot waveguide, the offset was set to 0 μm, ensuring direct contact between the hollow waveguide opening and the silicon slot coupler. Offsets along other axes were set to 0 μm, positioning the silicon waveguides at the hollow core center.

***Simulation results***

In the *z*-axis (top row, Fig. S8(c,d)), the optimal position for the slot coupler is located directly at the interface edge, eliminating the need for fine alignment. Meanwhile, the tapered-spike waveguide showed ripples in $|S_{21}|$ of approximately 1–2 dB, requiring adjustment of positions for a flat response. Consequently, coupling efficiency of the tapered spike becomes challenging once the waveguide is fixed, such as after packaging or at frequencies approaching 1 THz [S2, S3].

Notably, the EM-slot waveguide exhibited negligible performance degradation for offsets up to 20 μm along the *x*-axis (middle row, Fig. S8(c,d)). Even if the offset extends to 50 μm, $|S_{21}|$ showed a slight degradation of less than 1-dB in the high-frequency range. In contrast, the tapered spike revealed significant transmittance reduction, with multiple ripples even at a 10-μm offset. At 20 μm, $|S_{21}|$ decreased dramatically less than –10 dB, likely due to strong field interference within the narrow gap between the metallic wall and silicon core.

For out-of-plane (*y*-axis) misalignment (bottom row, Fig. S8(c,d)), the tapered spike again demonstrated greater sensitivity compared to the EM-slot waveguide. When the offset was 100 μm, the ripples appeared in the conventional waveguide, decreasing the 3-dB bandwidth. In the EM-slot waveguide suffered approximately 2 dB in the high-frequency ranges because of the mode mismatch. Nevertheless, the 3-dB bandwidth was maintained the whole WR-2.2 band.

Practically, alignment tolerances of approximately ±50 μm in both *x*- and *y*-directions are acceptable for the EM-slot waveguide, while the offset can be controlled under 20 μm using microscopes. Additionally, the EM-slot does not require the *z*-axis adjustment.

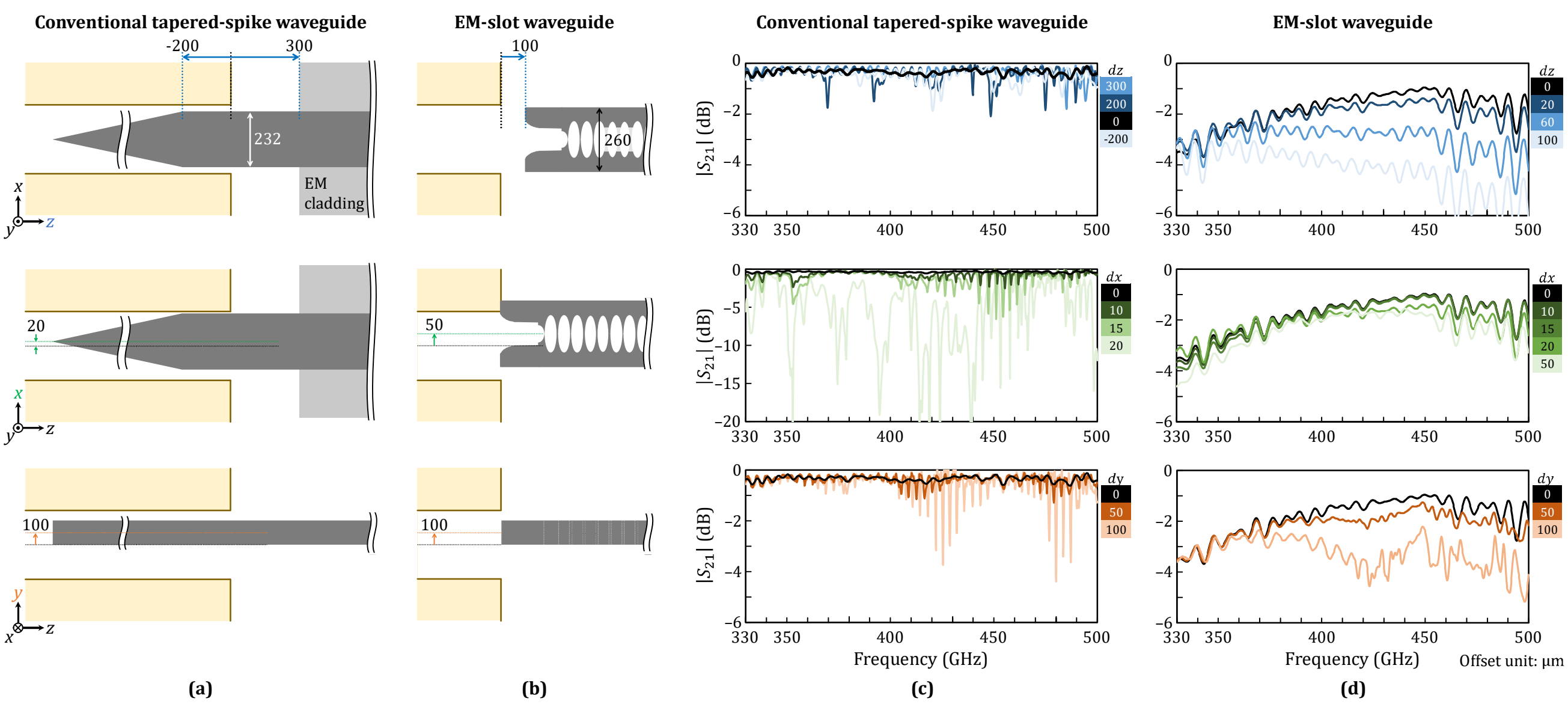

***Fig. S8*** *Simulation models for misalignment offsets of (a) conventional tapered-spike waveguide and (b) effective-medium (EM)-slot waveguide. Simulated transmission coefficients of the corresponding misalignment along three axes for (c) conventional tapered-spike waveguide and (d) EM-slot waveguide. TE-mode is excited at the hollow waveguide port. The misalignment in z-, x-, and y-axes are respectively presented in the top, middle, and bottom rows. The offset unit is in μm.*

ALT TEXT: Schematic of misalignment test in *x*-, *y*-, and *z*-directions. The conventional tapered-spike waveguide exhibits significant degradations for tolerances, thereby reducing 3-dB bandwidth, especially in *x*- and *z*-directions. On the other hand, the EM-slot waveguide shows insensitive in misalignment with reduction less than 1 dB in transmittance.

Overall, the EM-slot waveguide exhibits superior robustness to misalignment, making it a promising candidate for integration with hollow-waveguide interface and scalable packaging.

Here, we would like to highlight the risk associated with inserting a tapered spike into a metallic hollow waveguide. As shown in Fig. S9(a), scratches appear in the groove of the hollow core, leading to noticeable surface roughness. Consequently, the transmittance is significantly degraded, as illustrated in Fig. S9(b), making the waveguide unusable. By contrast, the proposed EM-slot waveguide avoids this issue because no silicon structure needs to be inserted into the hollow core for coupling. This spikeless coupling scheme therefore improves robustness and is advantageous for large-scale production of silicon waveguide-hollow waveguide systems.

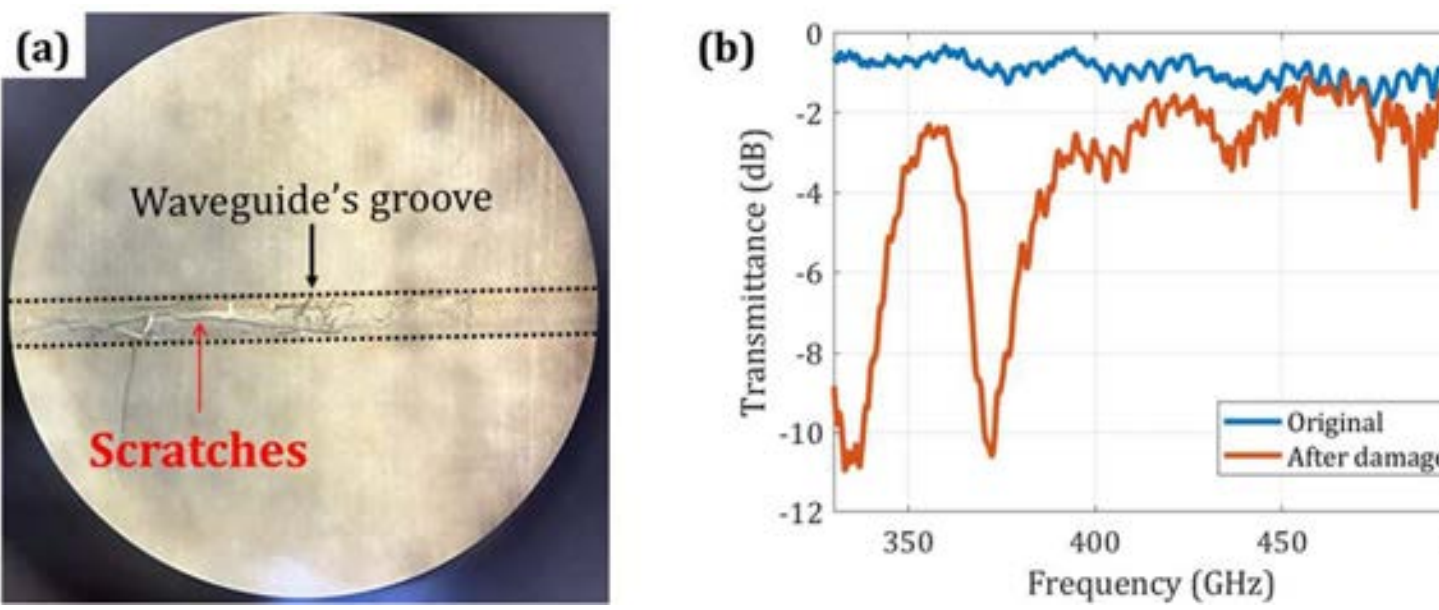

***Fig. S9*** *Risk of damage caused by inserting a tapered silicon spike into a metallic hollow waveguide. (a) Optical microscope image of scratches in the groove after breaking the taper inside. (b) Measured transmittance before (blue) and after (orange) the damage.*

ALT TEXT: Magnified image of the scratches on the surface of the groove of the metallic hollow waveguide caused by a tapered spike broken inside. This leads to substantial degradation of the transmittance.

## 3. Utilization of THz substrateless slot waveguide for interconnects

The substrateless slot waveguide with taper-free couplers enables robust coupling to hollow-waveguide-based devices without a fragile tapered spike, reducing the fine micro-machining required for conventional silicon waveguides. However, since the effective-medium channel consists of elliptical air holes, it is difficult to implement bending or furcating structures [S4-S6]. To address this limitation, simplifying the air holes by gradually transforming the hole size to a homogenous silicon channel can retain the slot interface while the channel can be flexibly designed with various shapes. Figure S10(a) illustrates the EM-slot waveguide with progressively reduced air-hole sizes, decreasing from 100% to 10% in steps of 10%. The coupling efficiency of this modified structure reaches approximately 90%, comparable to that of the original EM-slot waveguide as evidenced with an average difference of only 0.17 dB. The wavy coupling efficiency is likely due to the limited discrete changes of the effective refractive index ($n_{\mathrm{eff}}$) along the modification region. This can be solved by optimizing the sizes or increasing the number of reduced-size air holes, obtaining smoother change of $n_{\mathrm{eff}}$.

Consequently, the modified EM-slot waveguide can be potentially applied as a silicon platform with various functions, such as waveguides, multiplexers, and rotators. With the simple integration with THz active devices, the waveguide can be used for hybrid (electrical-THz-optical)-interconnection devices, such as an RTD (resonant tunneling diode) [S2, S7] or a uni-traveling carrier photodiode [S7, S8], as conceptually visualized by Fig. 1 of the main text.

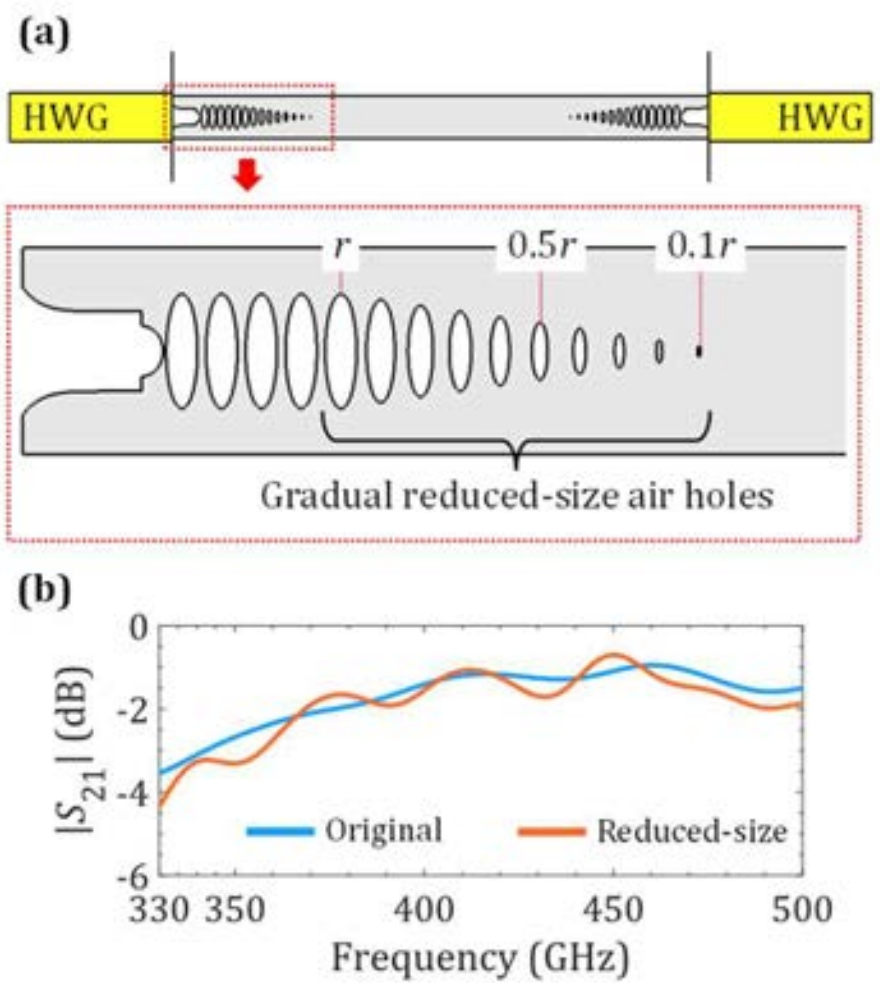

***Fig. S10*** *Substrateless slot waveguide transition. (a) EM-slot channel with gradual reduced-size air holes. The radius of each ellipse shrinks by 10% of the original ellipse's radius, from 100% to 10%. (b) Coupling efficiencies of the original EM-slot waveguide (blue) and the gradual reduced-size air-hole waveguide (orange).*

ALT TEXT: Schematic of the gradual reduced-size air holes of the EM-slot waveguide. The EM-slot channel can be transformed into a solid silicon core waveguide, where the radius of each ellipse shrinks by 10% of the original ellipse's radius, from 100% to 10%. Simulated transmittances show that the modified structure has comparable performance to the original structure.

### *4. Integration with electronic components*

The EM-slot waveguides are specifically designed for HWG -based devices, which are the industry de-facto standard for the sub-terahertz regime. To address broader integration with electronic components, we have adopted a hybrid interface strategy. For systems requiring interface with co-planar waveguides or microstrip lines [S9-S11], we utilize a tapered structure at one terminal to ensure robustness for electronic coupling, while the proposed EM-slot coupler is employed at the other for efficient hollow waveguide interfacing. Figure S11(a) shows the transition from a taper to solid core and subsequently to a slot coupler. This transmittance of the system is within –2.5 to –1.5 dB, which covers the whole WR-2.2 band, demonstrating the efficient coupling of this hybrid coupling scheme.

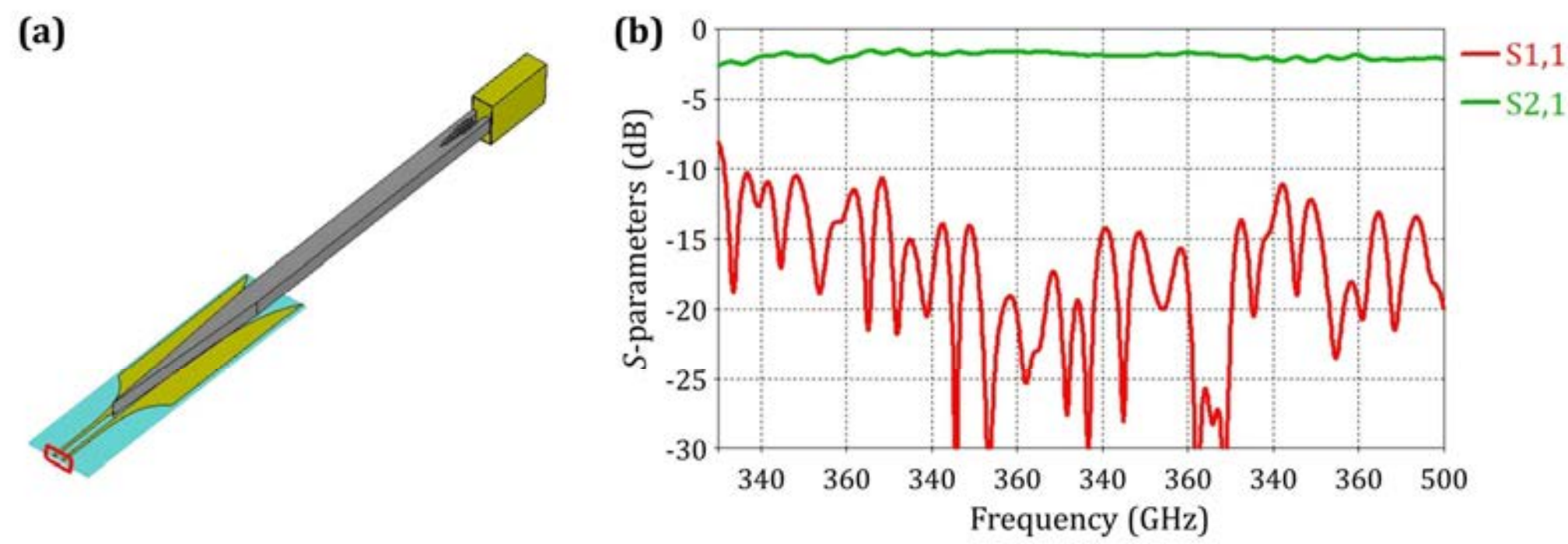

***Fig. S11*** *Silicon waveguide consisting of tapered spike and slot coupler. (a) Schematic and (b) simulated coupling efficiency as a function of frequency.*
ALT TEXT: Simulation model of a silicon waveguide with a modified slot coupler on one end and a tapered spike on the other. The coupler interfaces with a metallic waveguide, and the spike couples to a coplanar waveguide incorporating a Vivaldi antenna, achieving −2.5 to −1.5 dB transmittance across the WR-2.2 band.

Besides, we propose the integration of the EM-slot waveguide with a RTD chip, which is electronic device operating at terahertz band, coupling to a HWG, as described in Fig. S12(a). In [S12, S13], we presented a pioneering work of integrating RTD chips into silicon waveguides. Given the small footprint of the RTD (~1.5 μm$^2$), we utilized a modified Vivaldi antenna as a mode converter, as detailed in [S2], allowing the chip to be directly inserted into the EM channel. In Fig. S12(b), the simulation of E-field intensity shows that the waves are well confined within the air-hole array. Figure S12(c) show the transmittance of the system with a 3-dB bandwidth covering the whole WR-2.2 band and the maximum $|S_{21}|$ of –2.5 dB, demonstrating the broadband characteristic of the proposed coupling scheme for high-speed communications.

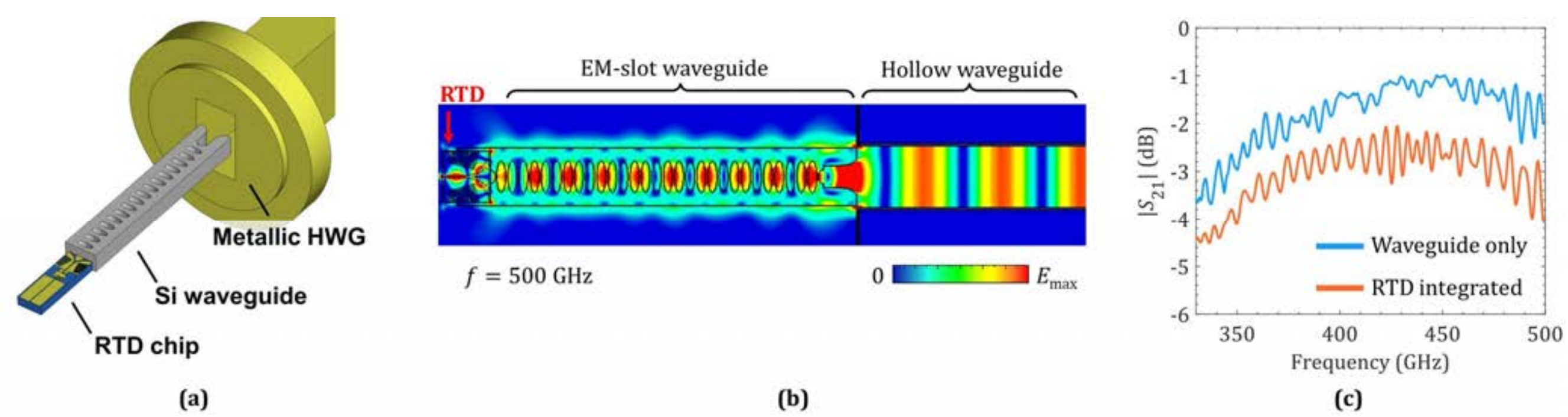

***Fig. S12*** *Integration of RTD chip with EM-slot waveguide for interconnection with hollow waveguide. (a) Schematic. (b) Simulated E-field distribution at 500 GHz. (c) Simulated coupling efficiency as a function of frequency.*
ALT TEXT: Simulation model of the RTD chip coupling to the EM-slot waveguide. The E-field distribution shows the waves are confined within the elliptical air holes, providing −4.5 to −2.5 dB transmittance across the WR-2.2 band.

## *5. Higher-order mode analysis*

### 1. Higher-Order Mode Analysis of EM-Channel

To evaluate modal distribution, we first analyzed the intrinsic modal properties of the elliptical air-hole channel without the slot couplers. The simulation was done by using a full wave simulator, CST Studio Suite 2025. Two waveguide ports were placed at two terminals of the EM channel, and the first five eigenmodes were excited, as shown in Figs. S13(a,b). In Fig. S13 (c), the simulated transmittance shows that:

- The fundamental mode dominates over the entire WR-2.2 and WR-1.5 bands.
- Higher-order modes exhibit significantly lower transmission and do not contribute appreciably within 330–600 GHz.
- The fundamental mode remains well confined and propagating up to ~750 GHz in the isolated EM channel.

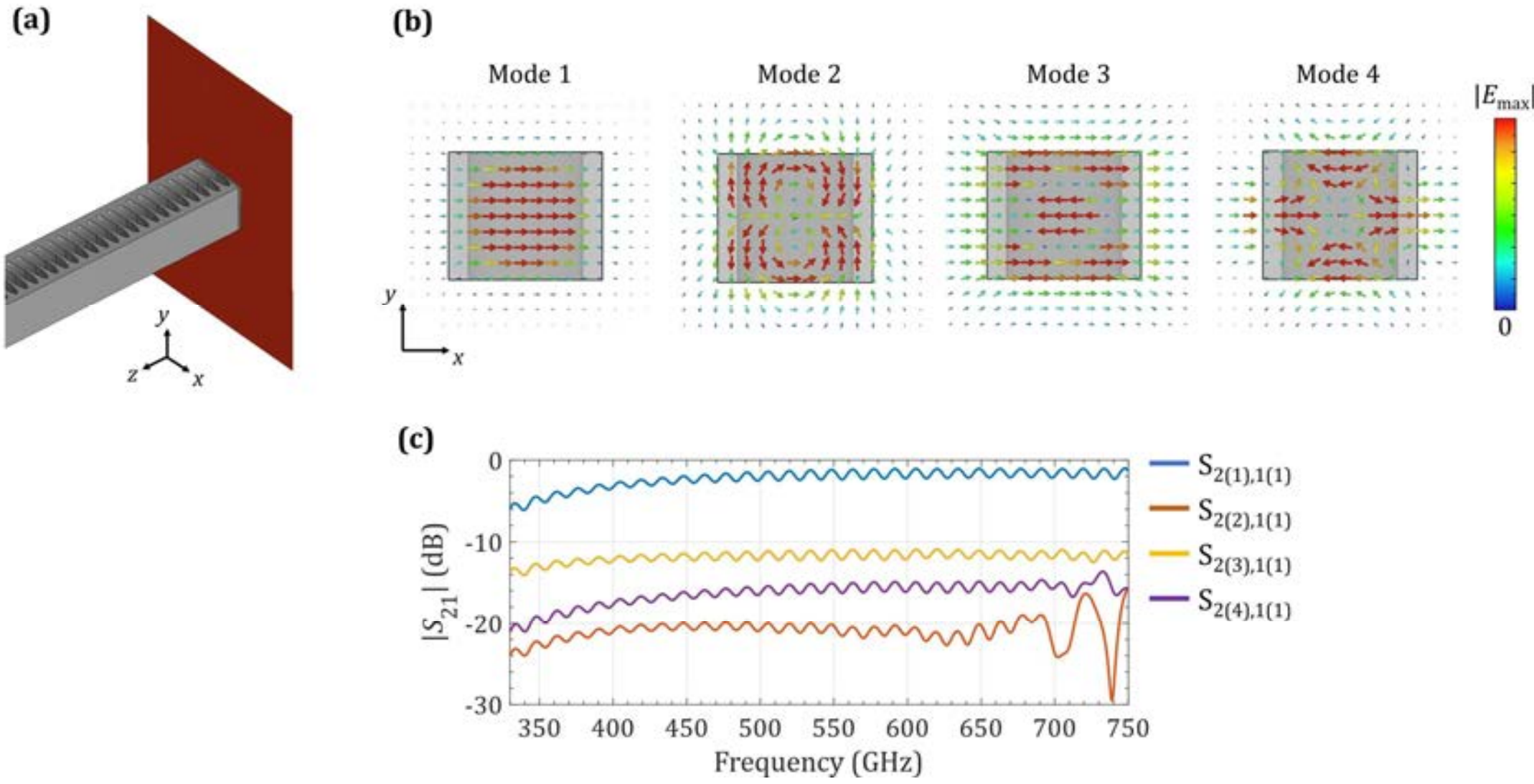

***Fig. S13*** *(a) Simulation model of the elliptical air-hole array. Waveguide port is fed at the terminal. (b) The first three modes generated at the waveguide port. (c)* $S_{21}$ *of the multiple modes.*

ALT TEXT: Simulation model of an EM-slot channel with waveguide ports at both ends. The first four modes are excited; the fundamental mode shows about −1 dB transmittance, while higher-order modes remain suppressed below −10 dB.

### 2. Mode Evolution After Introducing the Slot Coupler

After introducing the slot coupler, the complete structure was excited using the fundamental TE mode of the metallic hollow waveguide. As illustrated in Fig. S14(a), the excited field consistently corresponds to the fundamental slot mode within the 330–600 GHz communication band. The electric field is first strongly confined within the slot region and then gradually evolves into the EM-channel mode, indicating smooth modal transformation across the interface.

As the frequency increases beyond approximately 650 GHz, transmission degradation becomes noticeable. For example, the field at 700 GHz couples to the silicon beams of the slot coupler rather than confined inside the slot. This behavior is associated with the excitation of higher-order modes. In particular, the slot coupler was originally optimized for operation up to 500 GHz, which limits its high-frequency performance. This leads to excitation of higher-order modes and the decreased transmittance beyond the communication bandwidth, as shown in Fig. S14(b).

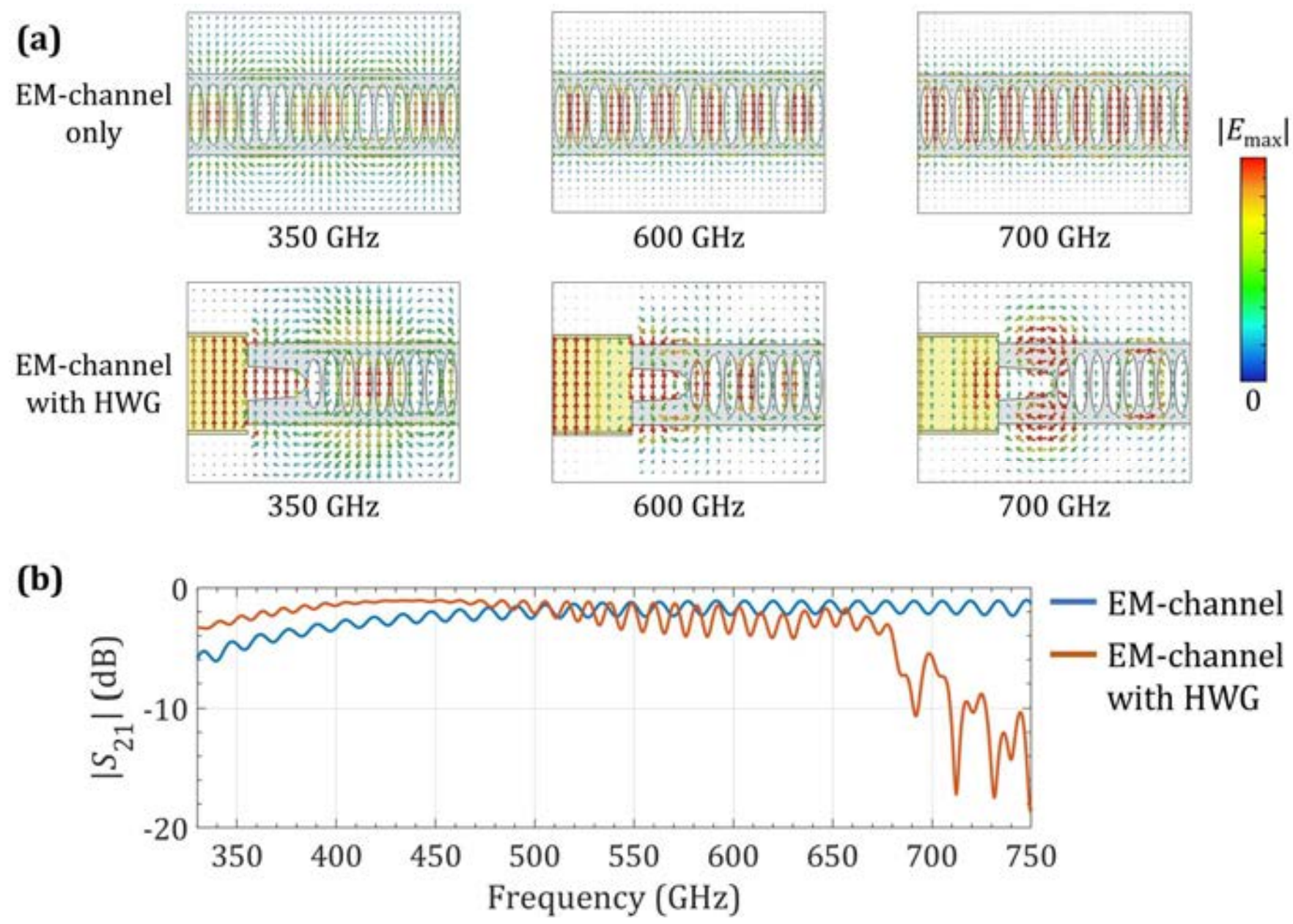

***Fig. S14*** *Observations of higher-order modes. (a) Vector E-field distributions of isolated EM-channel (top) and EM-slot waveguide coupled to metallic hollow waveguide (bottom). (b) Transmittance of the isolated EM-channel and EM-slot waveguide system.*

ALT TEXT: Vector fields of the EM-channel and EM-slot coupling to metallic hollow waveguide at different frequencies. At 350 GHz and 600 GHz, the waves are confined in the air-hole array in both waveguides. At 700 GHz, multimode is excited at the slot coupler of the EM-slot waveguide, thereby reducing the transmittance.

To evaluate the excited mode of the EM-slot waveguide coupling to metallic hollow waveguides, we calculate the overlap integral between the mode of the EM-slot waveguide and the isolated EM-channel as follows:

$$\eta = \frac{\iint E_{\mathrm{EM-slot}} \cdot E_{\mathrm{ref}}^{*}\, \mathrm{d}A}{\iint |E_{\mathrm{EM-slot}}|^{2}\, \mathrm{d}A \cdot \iint |E_{\mathrm{ref}}|^{2}\, \mathrm{d}A}$$

where $E_{\mathrm{EM-slot}}$ and $E_{\mathrm{ref}}$ are the $E$-fields of the EM-slot waveguide and isolated EM-channel, respectively. In Table S2, the calculated overlap integrals $\eta$ from 330–650 GHz is above 95%, demonstrating the dominant slot mode are excited in the waveguide. However, the overlap integral at 700 GHz significantly reduces 66.3%, showing that the slot mode gradually vanished, which is likely to the modal mismatch of the slot coupler.

***Table S2*** *Overlap integral between the mode of EM-slot waveguide and that of isolated EM-channel.*

| Frequency (GHz) | 350 | 400 | 450 | 500 | 550 | 600 | 650 | 700 | 750 |
|---|---|---|---|---|---|---|---|---|---|
| $\eta$ (%) | 98.5 | 98.9 | 98.7 | 98.0 | 98.8 | 98.6 | 95.5 | 66.3 | 71.3 |

ALT TEXT: Table data of overlap integral between the mode of EM-slot waveguide and reference waveguide. The efficiency remains more than 90% up to 650 GHz and significantly reduces to 66% beyond 700 GHz.